\documentclass{emulateapj}
 \usepackage{url}

\shorttitle{Magnetar central engine in GRBs}
\shortauthors{L\"{u} \& Zhang}
\slugcomment{}

\begin{document}

\title{A test of the millisecond magnetar central engine model of GRBs with Swift data}
\author{Hou-Jun L\"{u}, Bing Zhang}
\affil{Department of Physics and Astronomy, University of Nevada Las Vegas, Las Vegas, NV 89154, USA
\\lhj@physics.unlv.edu, zhang@physics.unlv.edu}

\begin{abstract}
A rapidly spinning, strongly magnetized neutron star (magnetar)
has been proposed as one possible candidate of the central
engine of gamma-ray bursts (GRBs). We systematically analyze
the Swift/XRT light curves of long GRBs detected before 2013
August, and characterize them into four categories based on how
likely they may harbor a magnetar central engine: Gold, Silver,
Aluminum, and Non-magnetar. We also independently analyze the
data of short GRBs with a putative magnetar central engine. We
then perform a statistical study of various properties of the
magnetar samples and the non-magnetar sample, and investigate
whether the data are consistent with the hypothesis that there
exist two types of central engines. By deriving the physical
parameters of the putative magnetars, we find that the
observations of the Gold and Silver samples are generally
consistent with the predictions of the magnetar model. For a
reasonable beaming factor for long GRBs, the derived magnetar
surface magnetic field $B_p$ and initial spin period $P_0$ fall
into the reasonable range. Magnetar winds in short GRBs, on the
other hand, are consistent with being isotropic. No GRB in the
magnetar sample has a beam-corrected total energy exceeding the
maximum energy budget defined by the initial spin energy of the
magnetar, while some non-magnetar GRBs do violate such a limit.
With beaming correction, on average the non-magnetar sample is
more energetic and luminous than the magnetar samples. Our
analysis hints that millisecond magnetars are likely operating
in a good fraction, but probably not all, GRBs.
\end{abstract}


\section{Introduction}
The central engine of gamma-ray bursts (GRBs) remains an open
question in GRB physics (Zhang 2011). Observations of GRB
prompt emission and early afterglow pose the following
constraints on a successful central engine model: (1) The
central engine must be able to power an outflow with an
extremely high energy and luminosity (e.g. Zhang \&
M\'{e}sz\'{a}ros, 2004; Meszaros 2006); (2) The ejecta must
have a low baryon loading, with energy per baryon exceeding 100
(e.g. Lithwick \& Sari, 2001; Liang et al. 2010); (3) The
central engine should be intermittent in nature to account for
the observed light curves with rapid variability (Fishman \&
Meagan 1995); (4) The engine should last for an extended period
of time to power delayed erratic X-ray flares (Burrows et al.
2005; Zhang et al. 2006) or long-lasting X-ray emission
followed by a sudden drop (i.e. ``internal plateau'', Troja et
al. 2007; Liang et al. 2007; Lyons et al. 2010); (5) Finally,
Fermi observations require that the central engine should be
strongly magnetized to launch a magnetically dominated outflow
at least for some GRBs (Zhang \& Pe'er 2009).

Two types of GRB central engine models have been discussed in
the literature (e.g. Kumar \& Zhang 2014 for a review). The
leading type of models invokes a hyper-accreting stellar-mass black hole
(e.g. Popham et al. 1999; Narayan et al. 2001; Lei et al.
2013), from which a relativistic jet is launched via
neutrino-anti-neutrino annihilation (Ruffert et al. 1997;
Popham et al. 1999; Chen \& Beloborodov 2007; Lei et al. 2009),
Blandford-Znajek mechanism (Blandford \& Znajek 1997; Lee et
al. 2000; Li 2000), or episodic magnetic bubble ejection from
the disk (Yuan \& Zhang 2012).

The second type of models invokes a rapidly spinning, strongly
magnetized neutron star dubbed a ``millisecond magnetar'' (Usov 1992;
Thompson 1994; Dai \& Lu 1998a; Wheeler et al. 2000; Zhang \&
M\'esz\'aros 2001; Metzger et al. 2008, 2011; Bucciantini et al.
2012). Within this scenario, the energy reservoir is the total
rotation energy of the millisecond magnetar, which
reads
\begin{equation}
E_{\rm rot} = \frac{1}{2} I \Omega_{0}^{2}
\simeq 2 \times 10^{52}~{\rm erg}~
M_{1.4} R_6^2 P_{0,-3}^{-2},
\label{Erot}
\end{equation}
where $I$ is the moment of inertia, $\Omega_0 = 2\pi/P_0$ is
the initial angular frequency of the neutron star, $M_{1.4} =
M/1.4M_\odot$, and the convention $Q = 10^x Q_x$ is adopted in
cgs units for all other parameters throughout the paper.

Assuming that the magnetar with initial spin period $P_0$ is
being spun down by a magnetic dipole with surface polar cap
magnetic field $B_p$, the spindown luminosity would evolve with
time as (Zhang \& M\'esz\'aros 2001)
\begin{eqnarray}
 L(t) &  = & L_0 \frac{1}{(1+t/\tau)^2} \nonumber \\
      & \simeq & \left\{
   \begin{array}{ll}
    L_0, & t \ll \tau, \\
    L_0 (t/\tau)^{-2}, & t \gg \tau.
   \end{array}
  \right.
\label{Lt}
\end{eqnarray}
where
\begin{equation}
 L_0 = 1.0 \times 10^{49}~{\rm erg~s^{-1}} (B_{p,15}^2 P_{0,-3}^{-4} R_6^6)
\label{L0}
\end{equation}
is the characteristic spindown luminosity, and
\begin{equation}
 \tau = 2.05 \times 10^3~{\rm s}~ (I_{45} B_{p,15}^{-2} P_{0,-3}^2 R_6^{-6})
\label{tau}
\end{equation}
is the characteristic spindown time scale.

The spin-down behavior of the magnetar can leave characteristic
imprints in the observed GRB emission. Dai \& Lu (1998a) first
proposed an energy injection model of millisecond pulsars to
interpret a rebrightening feature of the first optical
afterglow detected in GRB 970228. The required $B_p$ is $\sim
10^{13}$ G, not quite a magnetar strength. The prompt GRB
emission has to be attributed to additional physical processes,
e.g. magnetic dissipation in a differentially rotating neutron
star (Kluzniak \& Ruderman 1998) or strange quark star (Dai \&
Lu 1998b). Zhang \& M\'esz\'aros (2001) studied energy
injection from a central engine with a general luminosity law
$L(t) = L_0 (t/t_0)^{-q}$ (the magnetar injection corresponds
to $q=0$ for $t<\tau$ and $q=2$ for $t>\tau$), and pointed out
that besides the rebrightening feature discussed by Dai \& Lu
(1998a,b), for more typical magnetar parameters, one can have a
shallow decay phase followed by a normal decay phase in the
early afterglow of a GRB. Such a shallow decay phase (or
plateau) was later commonly observed in Swift early XRT light
curves (Zhang et al. 2006; Nousek et al. 2006; O'Brien et al.
2006; Liang et al. 2007). It can be readily interpreted as
energy injection from a millisecond magnetar central engine
(Zhang et al. 2006). An alternative energy injection model
invokes a short-duration central engine, which ejects materials
with a stratified Lorentz factor ($\Gamma$) profile. Energy is
gradually added to the blastwave as the blastwave is gradually
decelerated to progressively lower $\Gamma$ (Rees \&
M\'esz\'aros 1998; Sari \& M\'esz\'aros 2000; Uhm et al. 2012).
Both models can interpret the shallow decay phase of most X-ray
light curves.

A tie-breaker GRB was discovered in early 2007. GRB 070110
(Troja et al. 2007) showed an extended plateau with a near flat
light curve extending to over $10^4$ seconds before rapidly
falling off with a decay index $\alpha \sim 9$ (throughout the
paper the convention $F_\nu \propto t^{-\alpha} \nu^{-\beta}$
is adopted). Such a rapid decay cannot be accommodated in any
external shock model, so that the entire X-ray plateau emission
has to be attributed to internal dissipation of a central
engine wind. Such an ``internal plateau'' was later discovered
in several more GRBs (Liang et al. 2007; Lyons et al. 2010).
The near steady X-ray emission observed in GRB 070110 may not
be easy to interpret within a black hole central engine model,
but is a natural prediction of the magnetar central engine
model (Eq.(\ref{Lt}) when $t\ll \tau$). The rapid $t^{-9}$
decay near the end is not predicted in the magnetic dipole
radiation model. Troja et al. (2007) interpreted it as being
due to collapse of the magnetar to a black hole after loosing
centrifugal support\footnote{Such an interpretation recently
gains indirect support. Zhang (2014) suggested that such an
implosion in the GRB early afterglow phase should be
accompanied by a fast radio burst (FRB) (see also Falcke \&
Rezzolla 2013 for a proposal of more general supra-massive
neutron star implosions as the sources of FRBs), and tentative
detections of these FRBs following two GRBs may have been
detected (Bannister et al. 2012), roughly around the time
suggested by Zhang (2014).}. Interestingly, internal plateaus
are also discovered in a good fraction of short GRBs (Rowlinson
et al. 2010, 2013). Modeling various afterglow features for both
long and short GRBs within
the framework of the millisecond magnetar (or pulsar with weaker
magnetic field) central engine model has gained
growing attention (Dai et al. 2006; Gao \& Fan 2006; Fan \& Xu 2006;
Metzger et al. 2008, 2011; Dall'Osso et al. 2011; Fan et al. 2011;
Bucciantini et al. 2012;
Bernardini et al. 2013; Gompertz et al. 2013, 2014).
Numerical simulations of binary neutron star mergers indeed
show that a stable magnetar can survive if the initial masses
of the two neutron stars are small enough, which would power
a short gamma-ray burst (Giacomazzo \& Perna. 2013).

Even though evidence of a magnetar central engine is mounting,
it remains unclear whether the rich GRB data accumulated over
the years with the GRB mission Swift indeed statistically
requires the existence of (presumably) two types of central
engines. If indeed magnetars are operating in some GRBs while
hyper-accreting black holes are operating in others, do the
data show statistically significant differences between the two
samples? Do those GRBs that seem to have a magnetar signature
have physical parameters that are consistent with the
predictions of the magnetar central engine model?

This paper is to address these interesting questions through a
systematic analysis of the Swift X--Ray Telescope (XRT) data.
The XRT data reduction details and criteria for sample
selection are presented in \S 2. In \S 3, physical parameters
of the GRBs and the hypothetical magnetars are derived for all the
samples. A statistical comparison of the physical properties
between the magnetar samples and the non-magnetar sample are
presented in \S4, and conclusions are drawn in \S5 with some
discussion. Throughout the paper, a concordance cosmology with
parameters $H_0 = 71$ km s$^{-1}$ Mpc $^{-1}$, $\Omega_M=0.30$,
and $\Omega_{\Lambda}=0.70$ is adopted.

\section{Data reduction and sample selection criteria}

The XRT data are downloaded from the Swift data
archive\footnote{http://www.swift.ac.uk/archive/obs.php?burst=1}.
We developed a script to automatically download and maintain all
the XRT data on the local UNLV machine. The HEAsoft packages
{\em version 6.10}, including Xspec, Xselect, Ximage, and the
Swift data analysis tools, are used for the data reduction. An
IDL code was developed by the former group member B.-B. Zhang
to automatically process the XRT data for a given burst in any
user-specified time interval (see Zhang et al. 2007c for
details). We adopt this code with slight modifications to solve
the problem designed for this paper. The same IDL code was used
in several previous papers (Zhang et al. 2007c; Liang et al.
2007, 2008, 2009) of our group. More details about the
data reduction procedures can be found in Zhang et al. (2007c) and
Evans et al. (2009).

Our entire sample includes more than 750 GRBs observed between
2005 January and 2013 August, whose XRT data are all processed
with our data reduction tool. Since the magnetar signature
typically invokes a shallow decay phase (or plateau) followed
by a steeper decay segment (a normal decay for canonical light
curves, or a very steep decay for internal plateaus), our
attention is on those GRBs that show such a transition in the
X-ray light curves. We first identify such bursts by inspecting
their light curves. In order to grade their magnetar candidacy,
we next perform a temporal fit to the plateau behavior within a
time interval $(t_1, t_2)$, where $t_1$ is the beginning of the
plateau, while $t_2$ is the end of the segment after the
plateau break (either last observed data point if there is no
further break in the lightcurve, or the break time if a second
break appears). Since we are mostly interested in the behavior
around the break time $t_b$, the exact positions of $t_1$ and
$t_2$ do not matter much, so we pick them through visual
inspection of the light curves. We then fit the light curves
with a smooth broken power law
\begin{equation}
F = F_{0} \left[\left(\frac{t}{t_b}\right)^{\omega\alpha_1}+
\left(\frac{t}{t_b}\right)^{\omega\alpha_2}\right]^{-1/\omega},
\end{equation}
where $t_{b}$ is the break time, $F_b=F_0 \cdot 2^{-1/\omega}$
is the flux at the break time $t_b$,  $\alpha_1$ and $\alpha_2$
are decay indices before and after the break, respectively, and
$\omega$ describes the sharpness of the break. The larger the
$\omega$ parameter, the sharper the break.

An IDL routine named ``mpfitfun.pro'' is employed for our
fitting (Mor\'e 1977; Markwardt 2009). This routine performs a
Levenberg-Marquardt least-square fit to the data for a given
model to optimize the model parameters. After processing all
the data, we grade all long GRBs in our sample into four groups
(``Gold'', ``Silver'', ``Aluminum'', and ``non-magnetar'')
according to their likelihood of being powered by a magnetar
central engine.
\begin{itemize}
\item Gold: This sample is defined by those bursts that
    display an ``internal plateau''. These plateaus are
    followed by a decay slope steeper than 3, which is
    essentially impossible to interpret within the external
    shock models (Gao et al. 2013b)\footnote{The steepest
    decay slope in an external shock model is $2+\beta$
    (Kumar \& Panaitescu 2000), which is typically smaller
    than 3, and is defined by the high-latitude ``curvature
    effect'' emission from a conical outflow, even if the
    emission abruptly ceases.}. It demands a long-lasting
    central engine, and a near steady flux is consistent
    with emission from a spinning down magnetar. The rapid
    decay at the end of plateau may mark the implosion of
    the magnetar into a black hole (Troja et al. 2007;
    Zhang 2014). There are altogether only 9 robust cases
    identified in this Gold sample, 3 of which have
    redshift measurements. The light curves of these 9 GRBs
    together with the broken power-law fittings (red
    curves) are shown in Figure 1, and the fitting
    parameters are summarized in Table 1.
\item Silver: This sample includes GRBs with a shallow
    decay phase followed by a normal decay phase, and the
    pre- and post-break temporal and spectral properties
    are well consistent with the external forward shock
    model with energy injection of a magnetar as defined in
    Eq.(\ref{Lt}). Specifically, one requires two
    indepedent criteria to define this sample. First, the
    temporal and spectral properties of the afterglow after
    the break (the normal decay phase) should satisfy the
    ``closure relation'' of the external shock model (e.g.
    Zhang \& M\'esz\'aros 2004; Gao et al. 2013b), i.e.
\begin{equation}
\alpha_{2}=\cases{ \frac{3\beta}{2}=\frac{3(p-1)}{4},
            & $\nu_{m}<\nu<\nu_{c}$ (ISM) \cr
 \frac{3\beta+1}{2}=\frac{3p-1}{4},
            & $\nu_{m}<\nu<\nu_{c}$(Wind) \cr
\frac{3\beta-1}{2}=\frac{3p-2}{4},
            & $\nu>\nu_{c} $(ISM or Wind) \cr
            }
\end{equation}
    Here $\beta$ is the spectral index of the normal decay
    segment (which is X-ray photon index minus 1), and $p$
    is the electron's spectral distribution index. Second,
    the pre-break slope $\alpha_1$ should correspond to
    $q=0$, while the post-break slope $\alpha_2$ should
    correspond to $q=1$ (for a constant energy fireball,
    the scaling law is the same as $q=1$, Zhang \&
    M\'esz\'aros 2001), so according to Zhang et al. (2006)
    and Gao et al. (2013b), one should have
\begin{equation}
\alpha_{1}=\cases{ \frac{2\alpha_{2}-3}{3}, &
            $\nu_{m}<\nu<\nu_{c}$ (ISM) \cr
 \frac{2\alpha_{2}-1}{3}, & $\nu_{m}<\nu<\nu_{c}$(Wind) \cr
\frac{2\alpha_{2}-2}{3}, & $\nu>\nu_{c} $(ISM or Wind) \cr }
\label{a1-a2}
\end{equation}
    In our entire sample, 69 GRBs can be grouped into this
    Silver sample, with 33 having measured redshifts. The
    light curves with fitting curves are presented online
    at \url{http://grb.physics.unlv.edu/$\sim$lhj/Silver/},
    and the fitting results are reported in Table 1. Two
    examples (GRBs 060729, see also Grupe et al. 2007, and
    070306) are shown in Figure 2. Figure 3 shows all the
    GRBs in the $\alpha_1-\alpha_2$ plane, with three
    theoretically favored lines of the magnetar models
    (Eq.(\ref{a1-a2})) plotted. Those GRBs falling onto
    these lines (within error bars) and also satisfy the
    closure relations are identified as Silver sample GRBs
    (colored data points). In Fig.4 we present the
    distribution of electron spectral index $p$ derived
    from the Silver sample. It has a Gaussian distribution
    with a center value $p_{c}=2.51\pm0.04$. Figure 5 shows
    the distribution of Silver sample in the ($\alpha,
    \beta$)-plane combined with the closure relations for
    the models (ISM and wind medium).
\item Aluminum: Other GRBs with a shallow decay segment
    transiting to a steeper decay are included in the
    Aluminum sample. They either do not satisfy external
    shock closure relations in the post-break phase, or do
    not satisfy the $\alpha_1 - \alpha_2$ relations
    predicted in the magnetar external shock models. These
    are marked as grey points in Fig.3. Those GRBs that
    fall onto the three magnetar model lines but are still
    denoted as Aluminum are the ones that do not satisfy
    the closure relations in the post-break phase. On the
    other hand, since early magnetar spindown may not fully
    follow the simple dipole spindown law (e.g. Metzger et
    al. 2011), and since the observed X-ray emission may
    not come from the external forward shock emission (e.g.
    can be from external reverse shock, Dai 2004; Yu \& Dai
    2007, or from internal dissipation of the magnetar
    wind, Yu et al. 2010), these GRBs could be still
    powered by magnetars. We therefore still assign them as
    magnetar candidates, but with a lower grade. There are
    135 solid cases in the sample, 67 of which have
    redshift measurements. The light curves with fitting
    curves are presented online at
    \url{http://grb.physics.unlv.edu/$\sim$lhj/Aluminum/}.
    Two examples (GRBs 070420 and 080430) are presented
    in Fig.2.
\item Non-magnetar: All the other long GRBs we have
    analyzed are included in the non-magnetar sample. They
    either have a single power-law decay, or have erratic
    flares that prevent identifying a clear shallow decay
    phase, or present a rebrightening behavior, or the
    data are too poor to reach a robust conclusion. There
    are more than 400 GRBs in this group, 111 of which have
    redshift measurements. Strictly speaking, some of these
    GRBs may still host a magnetar central engine. We
    define these GRBs as ``non-magnetar'', simply because
    they do not present a clear magnetar signature. Two
    examples (GRBs 061007, see also Schady et al. 2007, Mundell
    et al. 2007, and 081028) are presented in the Fig. 2.
\end{itemize}

Finally, we also independently processed the X-ray data of short
GRBs that may harbor a magnetar central engine (cf. Rowlinson et
al 2013). We select the short GRBs that have measured
redshifts and high-quality X-ray data. The light curves with
fitting curves are presented online at
\url{http://grb.physics.unlv.edu/$\sim$lhj/SGRB/}.

\section{Derivations of the physical parameters}

Our purpose is to analyze and compare the physical properties
of GRBs with or without a magnetar signature. In this section,
we use data to derive relevant physical parameters. Redshift
measurements are crucial to derive the intrinsic parameters
(energy, luminosity, etc), so in the following we focus on
those GRBs with $z$ measurements only.

\subsection{Energetics, luminosity, and radiation efficiency}

The isotropic prompt $\gamma$-ray emission energy $E_{\rm
\gamma,iso}$ is usually derived from the observed fluence
$S_\gamma$ in the detector's energy band, and extrapolated to
the rest-frame $1-10^4$ keV using spectral parameters (the low-
and high- energy spectral indices $\hat\alpha$, $\hat\beta$,
and the peak energy $E_p$ for a standard ``Band-function'' fit,
Band et al. 1993) and through $k$-correction. However, since
the BAT energy band is narrow (15-150 keV), for most GRBs the
spectra can be only fit by a cutoff power law or a single power
law (Sakamoto et al. 2008, 2011). We therefore apply the
following procedure to estimate the Band spectral parameters:
(1) If a burst was also detected by {\em Fermi} GBM or {\em
Konus} Wind, we adopt the spectral parameters measured by those
instruments. (2) For those bursts that are not detected by
other instruments but can be fit with a cutoff power law model,
we adopt the derived $\hat\alpha$ and $E_p$
parameters\footnote{We note that usually the low-energy photon
index $\hat\alpha$ and $E_p$ are slightly different for the
cut-off power law and Band-function models (e.g. Sakamoto et
al. 2008, 2011), but the derived $E_{\rm \gamma,iso}$ only
shows a slight difference, which is ignored in our analysis.},
and assume a typical value of $\hat\beta = -2.3$. (3) For those
GRBs that can be only fit with a single power law, we have to a
derive $E_p$ using an empirical correlation between the
BAT-band photon index $\Gamma^{\rm BAT}$ and $E_p$ (e.g.
Sakamoto et al. 2009; Zhang et al. 2007b; Virgili et al. 2012;
L\"{u} et al. 2012). The typical parameters $\hat{\alpha}= -1$,
$\hat{\beta}=-2.3$ are adopted to perform the simulations. We
can then calculate the $E_{\rm \gamma,iso}$ according to
\begin{eqnarray}
E_{\rm \gamma,iso}&=&4\pi k D^{2}_{L} S_{\gamma} (1+z)^{-1}\nonumber \\
&=&1.3\times 10^{51}~ {\rm erg}~ k D^{2}_{28} (1+z)^{-1} S_{\gamma,-6}
\end{eqnarray}
where $z$ is the redshift, $D = 10^{28}~{\rm cm}~D_{28}$ is the
luminosity distance, and $k$ is the $k$-correction factor from
the observed band to $1-10^4$ keV in the burst rest frame (e.g.
Bloom et al. 2001).

Another important parameter is the isotropic kinetic energy
$E_{\rm K,iso}$ measured from the afterglow flux. This value is
increasing during the shallow decay phase, but becomes constant
during the normal decay phase (Zhang et al. 2007a). We follow
the method discussed in Zhang et al. (2007a) to calculate
$E_{\rm K,iso}$ during the normal decay phase using the X-ray
data. Noticing that fast-cooling is disfavored at this late
epoch, we derive several relevant cases. For $\nu>{\rm
max}(\nu_m,\nu_c)$, the afterglow flux expression does not depend
on the medium density, so the following expression (Zhang et al.
2007a) applies to both ISM and wind models\footnote{The
coefficients may be slightly different for the two ambient
medium models. Since in this regime one cannot differentiate
the two circumburst medium models, we universally adopt this
equation derived from the ISM model, keeping in mind that there
might be a factor of a few correction if the medium is
wind-like.}
\begin{eqnarray}
E_{\rm K,iso,52
} & = & \left[\frac{\nu F_\nu (\nu=10^{18}~{\rm Hz})}{5.2\times
10^{-14} ~{\rm ergs~s^{-1} ~cm^{-2}} }\right]^{4/(p+2)} \nonumber \\ &\times
&D_{28}^{8/(p+2)}(1+z)^{-1}
t_d^{(3p-2)/(p+2)}\nonumber \\
& \times & (1+Y)^{4/(p+2)} f_p^{-4/(p+2)}\epsilon_{B,-2}^{(2-p)/(p+2)} \nonumber \\
& \times &\epsilon_{e,-1}^{4(1-p)/(p+2)} \nu_{18}{^{2(p-2)/(p+2)}}.
\nonumber \\
\end{eqnarray}
For the $\nu_m < \nu < \nu_c$ ISM model, one has (Zhang et al.
2007a)
\begin{eqnarray}
E_{\rm K,iso,52} & = & \left[\frac{\nu F_\nu (\nu=10^{18}~{\rm Hz})}{6.5\times
10^{-13} ~{\rm ergs~s^{-1} ~cm^{-2}} }\right]^{4/(p+3)} \nonumber  \\& \times &
D_{28}^{8/(p+3)}(1+z)^{-1}
 t_d^{3(p-1)/(p+3)}\nonumber \\
& \times &f_p^{-4/(p+3)} \epsilon_{B,-2}^{-(p+1)/(p+3)}
\epsilon_{e,-1}^{4(1-p)/(p+3)} \nonumber \\ & \times & n^{-2/(p+3)}
\nu_{18}{^{2(p-3)/(p+3)}}.
\nonumber \\
\end{eqnarray}
For the $\nu_m < \nu < \nu_c$ wind model, one has (Gao et al. 2013b)
\begin{eqnarray}
\nu_{m}=5.5\times10^{11}Hz (\frac{p-2}{p-1})^{2}(1+z)^{1/2}\epsilon^{1/2}_{B,-2}
\epsilon^{2}_{e,-1}E^{1/2}_{\rm K,iso,52}t^{-3/2}_{d},\nonumber \\
\end{eqnarray}
\begin{eqnarray}
\nu_{c}=4.7\times10^{18}Hz(1+z)^{-3/2}A^{-2}_{\ast,-1}
\epsilon^{-3/2}_{B,-2}E^{1/2}_{\rm K,iso,52}t^{1/2}_{d}, \nonumber \\
\end{eqnarray}
\begin{eqnarray}
F_{\nu,max}=5.7\times10^{2}\mu Jy(1+z)^{3/2}A_{\ast,-1}
\epsilon^{1/2}_{B,-2}D^{-2}_{28}E^{1/2}_{\rm K,iso,52}t^{-1/2}_{d}, \nonumber \\
\end{eqnarray}
so that
\begin{eqnarray}
\nu F_{\nu}(\nu=10^{18}Hz)&=&\nu F_{\nu,max}(\frac{\nu}{\nu_{m}})^{-(p-1)/2} \nonumber \\
&=& F_{\nu,max}\nu^{(3-p)/2}\nu^{(p-1)/2}_{m} \nonumber \\
&=& 7.4\times10^{-14}~{\rm erg~cm^{-2}~s^{-1}}~ \nonumber \\
&\times & D^{-2}_{28} (1+z)^{(p+5)/4}
A_{\ast,-1}f_{p}\epsilon^{(p+1)/4}_{B,-2}\epsilon^{p-1}_{e,-1}\nonumber \\
&\times &E^{(p+1)/4}_{\rm K,iso,52}t^{(1-3p)/4}_{d} \nu^{(3-p)/2}_{18},
\end{eqnarray}
and
\begin{eqnarray}
E_{\rm K,iso,52} & = & \left[\frac{\nu F_\nu (\nu=10^{18}~{\rm Hz})}{7.4\times
10^{-14} ~{\rm ergs~s^{-1} ~cm^{-2}} }\right]^{4/(p+1)} \nonumber  \\& \times &
D_{28}^{8/(p+1)}(1+z)^{-(p+5)/(p+1)}
 t_d^{(3p-1)/(p+1)}\nonumber \\
& \times &f_p^{-4/(p+1)} \epsilon_{B,-2}^{-1}
\epsilon_{e,-1}^{4(1-p)/(p+1)} \nonumber \\ & \times & A_{\ast,-1}^{-4/(p+1)}
\nu_{18}{^{2(p-3)/(p+1)}}.
\end{eqnarray}
Here $\nu f_\nu(\nu=10^{18}{\rm Hz})$ is the energy flux at
$10^{18}$ Hz (in units of ${\rm ergs~s^{-1} ~cm^{-2}}$), $n$ is
the density of the constant ambient medium, $A_*$ is the
stellar wind parameter, $t_d$ is the time in the observer frame
in days, and $Y$ is the Compton parameter. The electron
spectral index $p$ and the spectral index $\beta$ are connected
through
\begin{equation}
p=\cases{ 2\beta+1,
            & $\nu_{m}<\nu<\nu_{c}$ \cr
 2\beta,
            & $\nu>\nu_{c}$, \cr
            }
\end{equation}
and $f_p$ is a function of the power law distribution
index $p$ (Zhang et al. 2007a)
\begin{equation}
f_{p}\sim6.73\left(\frac{p-2}{p-1}\right)^{p-1} (3.3\times 10^{-6})^{(p-2.3)/2}
\end{equation}
In our calculations, the microphysics parameters of the shock
are assigned to standard values dervied from observations (e.g.
Panaitescu \& Kumar 2002; Yost et al. 2003): $\epsilon_{e}$=0.1
and $\epsilon_{B}=0.01$. The Compton parameter is assigned to a
typical value $Y=1$.

After deriving the break time $t_b$ through light curve fitting,
we derive the break time luminosity as
\begin{equation}
L_b = 4\pi D^2 F_b,
\end{equation}
where $F_b$ is the X-ray flux at $t_b$. Since the XRT band is
narrow, no $k$-correction is possible to calculate $L_b$.

A jet break was detected in some GRBs in our sample.
For these GRBs, we correct all the isotropic values to the
beaming-corrected values by multiplying the values by the
beaming correction factor (Frail et al. 2001)
\begin{equation}
f_b = 1-\cos \theta_j \simeq (1/2) \theta_j^2,
\end{equation}
i.e. $E_\gamma=E_{\rm \gamma,iso} f_b$,
and $E_{\rm K} = E_{\rm K,iso} f_b$. The jet
angle information was searched from the literature (e.g. Liang
et al. 2008; Racusin et al. 2009; Lu et al. 2012; Nemmen et al.
2012), which is collected in Table 2.

The GRB radiation efficiency is defined as (Lloyd-Ronning \&
Zhang 2004)
\begin{equation}
\eta_{\gamma} = \frac{E_{\rm \gamma,iso}}{E_{\rm \gamma,iso}+E_{\rm K,iso}}
= \frac{E_\gamma}{E_\gamma+E_{\rm K}}.
\end{equation}
Since $E_{\rm K,iso}$ (and $E_{\rm K}$) are increasing
functions of time during the shallow decay phase,
$\eta_{\gamma}$ is different when $E_{\rm K,iso}$ ($E_{\rm K}$)
at different epochs are adopted. Following Zhang et al.
(2007a), we take a typical blastwave deceleration $t_{dec}$ and
the end of the shallow decay phase $t_b$ to calculate the
radiative efficiencies. Within the framework of the magnetar
central engine model, the two efficiencies carry different
physical meanings: $\eta_{\gamma}(t_{dec})$ denotes the
efficiency of dissipating the magnetar wind energy during the
prompt emission phase, while $\eta_{\gamma}(t_b)$ denotes the
total efficiency of converting the spindown energy of a
magnetar to $\gamma$-ray radiation.

\subsection{Magnetar parameters}

For a magnetar undergoing dipolar spindown, two important
magnetar parameters, i.e. the initial spin period $P_0$ and the
surface polar cap magnetic field $B_p$, can be solved by the
characteristic luminosity $L_0$ (Eq.(\ref{L0})) and the
spindown time scale $\tau$ (Eq.(\ref{tau})).

The spindown time scale can be generally identified as the
observed break time, i.e.
\begin{equation}
\tau = t_b/(1+z).
\label{tau=tb}
\end{equation}
One caution is that $\tau$ can be shorter than $t_b/(1+z)$ if the
magnetar is supra-massive, and collapses to a black hole before
it is significantly spun down. On the other hand, the angular
velocity of the magnetar does not change significantly until
reaching the characteristic spindown time scale, so that the
collapse of the supra-massive magnetar, if indeed happens,
would likely happen at or after $\tau$. In our analysis, we
will adopt Eq.(\ref{tau=tb}) throughout.

The characteristic spindown luminosity should generally include
two terms:
\begin{equation}
 L_0 = L_{\rm X} + L_{\rm K} = (L_{\rm X,iso} + L_{\rm K,iso}) f_b,
\label{L0-2}
\end{equation}
where $L_{\rm X,iso}$ is the X-ray luminosity due to internal
dissipation of the magnetar wind, which is the observed X-ray
luminosity of the internal plateau (for external plateaus, one
can only derive an upper limit), and
\begin{equation}
L_{\rm K,iso} = E_{\rm K,iso} (1+z) / t_b
\end{equation}
is the kinetic luminosity that is injected into the blastwave
during the energy injection phase. It depends on the isotropic
kinetic energy $E_{\rm K,iso}$ after the injection phase is
over, which can be derived from afterglow modeling discussed
above. For the Gold sample, the $L_{\rm X,iso}$ component
dominates, while for Silver and Aluminum samples, the $L_{\rm
K,iso}$ component dominates. In any case, both components
should exist and contribute to the observed flux (Zhang 2014).
One can also define an X-ray efficiency to define the radiative
efficiency for a magnetar to convert its spindown energy to
radiation, i.e.
\begin{equation}
\eta_{\rm X} = \frac{L_{\rm X}}{L_{\rm X}+L_{\rm K}} =
\frac{L_{\rm X,iso}}{L_{\rm X,iso}+L_{\rm K,iso}}.
\label{etaX}
\end{equation}

In our analysis, we try to calculate both $L_{\rm X,iso}$ and
$L_{\rm K,iso}$ from the data. For the Gold sample GRBs that
show internal plateaus, $L_{\rm X,iso}$ can be readily
measured. For the cases where the internal plateau lands on an
external shock component (e.g. Troja et al. 2007), $L_{\rm
K,iso}$ can be also derived by modeling the late X-ray
afterglow in the normal decay phase. For the Gold sample cases
where no late external shock component is available, one can
only set up an upper limit on $L_{\rm K,iso}$. For Silver and
Aluminum samples, the internal plateau component is not
detectable. Through simulations, we find that the external
shock component would not be significantly modified if the
internal plateau flux is below 50\% of the observed external
shock flux. Therefore for all the Silver and Aluminum sample
GRBs, we place an upper limit of $L_{\rm X,iso}$ as 50\% of the
observed X-ray flux.

\section{Results}

\subsection{Magnetar parameters and collimation}

We derive magnetar parameters ($P_0$ and $B_p$) of the Gold,
Silver and Aluminum samples using Eqs.(\ref{L0}), (\ref{tau}),
and (\ref{L0-2})\footnote{Strictly speaking, these magnetar
parameters are the ones after prompt emission is over, since
only $L_{\rm X}$ and $L_{\rm K}$ are used to derive them. The
GRB prompt emission presumably also consumed spin energy and
magnetic energy of the magnetar, so the true initial spin
period can be somewhat smaller than $P_0$, and the true initial
(effective) dipole magnetic field at the pole can be somewhat
larger than $B_p$).}. First, we assume that the magnetar wind
is isotropic, so that $f_b = 1$. The derived $P_0$, $B_p$ are
presented in Table 2 and Fig.6a. Most ``magnetars'' have $B_p$
below $10^{15}$ G, some even have $B_p$ below $10^{13}$ G,
which are not considered as magnetars. More problematically,
most derived $P_0$'s are much shorter than 1 ms. This directly
conflicts with the break-up limit of a neutron star, which is
about 0.96 ms (Lattimer \& Prakash 2004). This suggests that
the isotropic assumption for these long GRB magnetar winds is
not correct. We then introduce the beaming factor $f_b$ for
each GRB. If $\theta_j$ is measured, we simply adopt the value.
Otherwise, we choose $\theta_j=5^{\rm o}$, a typical jet
opening angle for bright long GRBs (Frail et al. 2001; Liang et
al. 2008). Very interestingly, after such a correction, all the
data points of Gold and Silver sample GRBs fall into the
expected region in the $P_0-B_p$ plot (Fig.6b).
Also the additional conditions imposed by the causality
argument (i.e. that the speed of sound on the neutron star
cannot exceed the speed of light, Lattimer et al. 1990, and
Eqs.(9) and (10) of Rowlinson et al. (2010)) are satisfied for
all GRBs in all three (Gold, Silver and Aluminum) magnetar
samples, if one assumes $M=1.4M_{\odot}$. All these suggest
that the long GRB magnetar winds are likely collimated. Some
Aluminum sample GRBs are still to the left of the allowed
region (with $P_0$ shorter than the break-up limit). This may
suggest that those Aluminum sample bursts are not powered by
magnetars, or are powered by magnetars with even narrower jets.

Very interestingly, the magnetar properties of short GRBs
derived under the isotropic assumption actually lie reasonably
in the allowed region (Fig.6a, blue dots). After jet correction
for long GRB magnetars (but keep short GRB magnetar wind
isotropic), the derived magnetar parameters are well mixed in
the same region.  This suggests that the isotropic assumption
for short GRBs is reasonably good. This is understandable
within the framework of the progenitor models of GRBs. Short
GRBs are believed to be powered by mergers of NS-NS or NS-BH
systems (Paczynski 1986; Eichler et al. 1989; Paczynski 1991;
Narayan et al. 1992). During the merger process, only a small
amount of materials are launched (Freiburghaus et al. 1999;
Rezzolla et al. 2010; Hotokezaka et al. 2013). A millisecond
magnetar is expected to launch a near isotropic wind. This
wind, instead of being collimated by the ejecta (e.g.
Bucciantini et al. 2012), would simply push the ejecta behind
and accelerate the ejecta and make a bright electromagnetic
signal in the equatoral directions (Fan \& Xu 2006; Zhang 2013;
Gao et al. 2013; Yu et al. 2013; Metzger \& Piro 2013). In the jet
direction, the magnetar wind emission is not enhanced by the
beaming effect, so that one can infer correct magnetar
parameters assuming an isotropic wind. For long GRBs, on the
other hand, jets are believed to be launched from collapsing
massive stars (Woosley 1993; MacFadyen \& Woosley 1999). The
initially near isotropic magnetar wind is expected to be soon
collimated by the stellar envelope to a small solid angle
(Bucciantini et al. 2008).

\subsection{Statistical properties and correlations of other parameters}

Figure 7 shows the correlations of $L_{b}-E_{\rm \gamma,iso}$ and
$L_{b}-t_{b}$ for the entire sample. As shown in Fig.7a, a
higher isotropic $\gamma$-ray energy generally has a higher
X-ray break luminosity. For the Gold and Silver samples, a
Spearman correlation analysis gives a dependence
\begin{equation}
\log L_{b,49}=(1.48 \pm 0.17) \log E_{\rm \gamma,iso,52}+(2.56 \pm 0.75),
\end{equation}
with a correlation coefficient $r=0.83$, and a chance probability
$p<0.001$. Adding the Aluminum sample only
slightly worsens the correlation
($\log L_{b,49}=(1.02 \pm 0.10) \log E_{\rm \gamma,iso,52}+(2.64 \pm 2.04)$,
with $r=0.72$ and $p<0.001$).
Such a correlation is expected, which may be caused by a combination of
intrinsic (a more energetic magnetar gives more significant
contribution to both prompt emission and afterglow) and geometric
effects (a narrower jet would enhance both prompt emission and
afterglow).

Figure 7b presents an anti-correlation between $L_b$ and $t_b$
(Dainotti et al. 2010). Our Gold + Silver sample gives
\begin{equation}
\log L_{b,49}=(-1.83 \pm 0.20) \log t_{b,3}+(0.2 \pm 0.18)
\end{equation}
with $r=0.84$ and $p<0.001$. Adding the Aluminum sample only
slightly worsens the correlation
($\log L_{b,49}=(-1.29 \pm 0.15) \log t_{b,3}-(0.43 \pm 0.14)$
with $r=0.66$ and $p<0.001$). Such an anti-correlation is consistent
with the prediction of the magnetar model: Given a quasi-universal
magnetar total spin energy, a higher magnetic field would power a
brighter plateau with a shorter duration, or vice versa
(see also Xu \& Huang 2012).

In Fig.8, we compare the inferred $E_{\gamma}+E_{\rm K}$ with
the total rotation energy $E_{\rm rot}$ (Eq.(\ref{Erot})) of
the millisecond magnetar. It is found that the GRBs are
generally above and not too far above the $E_{\rm rot} =
E_{\gamma} + E_{\rm K}$ line. This is consistent with the
magnetar hypothesis, namely, all the emission energy ultimately
comes from the spin energy of the magnetar. Figure 8a includes
all the GRBs in the Gold/Silver/Aluminum samples, with
$\theta_j = 5^{\rm o}$ assumed if the jet angle is not
measured. Figure 8b presents those GRBs with jet measurements
only. Essentially the same conclusion is reached.

A very interesting question is whether there are noticeable
differences between the magnetar and non-magnetar samples. One
potential discriminator would be the total energetics of the
GRBs. While the magnetar model predicts a maximum value of the
total energy (Eq.(\ref{Erot})), the black hole model is not
subject to such a limit. In Fig.9 we make some comparisons. The
first three panels compare the histograms of the isotropic
energies ($E_{\rm \gamma,iso}$, $E_{\rm K,iso}$, and $E_{\rm
\gamma,iso}+E_{\rm K,iso}$) of the magnetar and non-magnetar
samples. For the magnetar sample, we in one case includes the
most secure (Gold + Silver) sample only (blue hatched), and in
another case includes all magnetar candidates (Gold + Silver +
Aluminum) (red solid). The non-magnetar sample is marked in
grey. The best Gaussian fits to the three samples are presented
as blue, red, and black dotted curves, respectively. The center
values of all the fits are presented in Table 3. It is found
that without jet correction, the isotropic values of the
magnetar and non-magnetar samples are not significantly
different.

Next, we introduce beaming correction, and replot the
histograms of the jet-corrected energies of the magnetar and
non-magnetar samples. The results are presented in the later
three panels in Fig.9. One can see a clear distinction between
the robust magnetar sample (Gold + Silver) and the non-magnetar
sample. For the total energy ($E_{\gamma} + E_{\rm K}$), while
the former peaks around 50.62 erg, the latter peaks around
51.81 erg. More interestingly, all the Gold+Silver magnetar
sample GRBs have a total energy smaller than the limit set by
the spin energy (Eq.(\ref{Erot})), while for some non-magnetar
sample GRBs, this upper limit is exceeded. The results are
generally consistent with the hypothesis that two types of GRB
central engines can both power GRBs.

In Fig.10a and Fig.10b, we compare $E_{\rm \gamma,iso}$ and
$E_{\rm K,iso}$ for the magnetar and non-magnetar samples. The
kinetic energy of the blastwave $E_{\rm K,iso}$ is evaluated at
$t_b$ for Fig.10a, and at $t_{dec}$ for Fig.10b (similar to
Zhang et al. 2007a). It is interesting to see that at
$t_{dec}$, the magnetar central engine tends to power more
efficient GRBs (due to the initial small $E_{\rm K}$ value)
than the black hole central engine. It is interesting to see
after the energy injection phase (at $t_b$), the $\gamma$-ray
efficiencies of magnetar and non-magnetar samples are no longer
significantly different. The same conclusion is also manifested
in Fig.11a and 11b, where we plot the histograms of
$\eta_{\gamma}$ for different samples.

If one accepts that millisecond magnetars are powering some
GRBs, it would be interesting to constrain the internal energy
dissipation efficiency $\eta_{\rm X}$ (Eq.(\ref{etaX})) from
the data. In both Fig.7a and Fig.7b, it is found that the Gold
sample GRBs have a relatively large $L_b$ value. This is
generally consistent with the expectation that a larger
$\eta_{\rm X}$ would give rise to an internal plateau (Zhang
2014). In Fig.10c, we compare $L_{\rm K,iso}$ and $L_{\rm
X,iso}$. It indeed shows that the Gold sample GRBs have a much
higher $\eta_{\rm X}$ than other GRBs. On the other hand, it is
curious to ask why there is a gap in this phase space. It
appears that some magnetars are particularly efficient to
dissipate the magnetar wind energy, while most magnetars are
not. Plotting the histograms of $\eta_{\rm X}$ (Fig.11c), it
looks indeed like a bimodal distribution of $\eta_{\rm X}$,
even though this second high $\eta_{\rm X}$ component is not
significant enough. In Fig.12, we present the scatter plots of
$\eta_{\rm X}$ against other parameters, including $\eta_\gamma
(t_b)$, $E_{\rm \gamma,iso}$, $E_{\rm K,iso}$, and $E_{\rm
rot}$. In all cases, the Gold sample (the ones with very high
$\eta_{\rm X}$) tend to stick out and emerge as a separate
population.

\section{Conclusions and Discussion}

In order to address whether (at least) some GRBs might have a
magnetar central engine, we have systematically analysed the
X-ray data of all the {\em Swift} GRBs ($\sim 750$) detected
before August 2013. By applying some criteria to judge how likely
a GRB might harbor a millisecond magnetar central engine, we
characterized long GRBs into several samples: Gold, Silver, and
Aluminum magnetar samples, as well as the non-magnetar samples.
For comparison, we also independently processed the data of
short GRBs that might have a magnetar central engine (Rowlinson
et al. 2010, 2013). By deriving the basic magnetar parameters
$P_0$ and $B_p$ from the data, we are able to reach two interesting
conclusions.

First, it seems that at least for the Gold and Silver sample GRBs,
the derived properties seem to be consistent with the expectations
of the magnetar central engine model. The consistency includes the
following: 1. After beaming correction, the derived $P_0$ and $B_p$
seem to fall into the reasonable range expected in the magnetar
central engine model; 2. The $L_b - t_b$ anti-correlation seems
to be consistent with the hypothesis that there is a quasi-universal
energy budget defined by the spin energy of the magnetars
(Eq.(\ref{Erot})); 3. The sum of $E_{\rm \gamma}$ and $E_{\rm K}$
is generally smaller than $E_{\rm rot}$, the total energy budget
of a magnetar; 4. Most importantly, it seems that the magnetar and
non-magnetar samples are different.
The robust magnetar sample (Gold + Silver) GRBs
all have a beaming-corrected energy smaller than the maximum energy
allowed by a magnetar, i.e. $E_{\rm rot,max} \sim 2\times 10^{52}$
erg. The non-magnetar sample, on the other hand, can exceed this
limit. The two samples have two distinct distributions in $E_\gamma$,
$E_{\rm K}$, and $(E_{\gamma}+E_{\rm K})$, suggesting that they
may be powered by different central engines.

Second, both long and short GRBs can be powered by a millisecond
magnetar central engine. The characteristic magnetar signature,
an internal plateau, is found in both long and short GRBs,
suggesting that different progenitors (both massive star core
collapses and compact star mergers) can produce a millisecond,
probably supra-massive magnetar as the central engine. The data
is consistent that a long GRB magnetar wind is collimated, while
a short GRB magnetar wind is essentially isotropic. All these
have profound implications in several related fields in high-energy,
transient astronomy. For example, if the recently discovered fast
radio bursts (FRBs, Lorimer et al. 2007; Thornton et al. 2013)
are indeed produced when a supra-massive neutron star collapses
into a black hole (Falcke \& Rezzolla 2014; Zhang 2014), our
analysis suggests that such supra-massive neutron stars very
likely do exist in GRBs, and that the FRB/GRB association suggested by
Zhang (2014) should be quite common, probably up to near half of
the entire GRB population. This is higher than the rate of plausible
detections made by Bannister et al. (2012), but that low detection
rate (2 out of 9 GRBs, Bannister et al. 2012) may be due to the
sensitivity limit of the Parkes 12 m telescope they have used.
A rapid-slewing larger radio telescope would be able to detect
more FRB/GRB associations, which would open a new window to study
cosmology (Deng \& Zhang 2014) and conduct cosmography
(Gao et al. 2014). For another example, the conclusion
that short GRBs can be powered by a millisecond magnetar with a
near isotropic magnetar wind would give rise to relatively
bright, early electromagnetic counterparts of gravitational wave
bursts due to NS-NS mergers (Zhang 2013; Gao et al. 2013; Yu et
al. 2013; Metzger \& Piro 2013; Fan et al. 2013), which gives
promising prospects of detecting electromagnetic counterparts of
gravitational wave signals in the Advanced LIGO/Virgo era.

Our analysis also poses some curious questions. One is regarding the
magnetar dissipation efficiency $\eta_{\rm X}$. The results seem to suggest
that some magnetars are efficient in dissipating their magnetar wind
energy to X-ray radiation, while most others are not. A straightforward
inference would be that there might be a dichotomy within the magnetar
central engines. A more plausible scenario would be that some (or
probably) most normal plateaus (those followed by normal decays)
could be also dominated by internal dissipation emission (e.g. Ghisellini
et al. 2007; Kumar et al. 2008a,b). They
are not identified as internal plateaus because their post-break
decay is not steep enough. Physically they may be stable magnetars
or supra-massive magnetars with a much later collapsing time, so that
the collapsing signature (very steep decay) is not detected. If so,
the $\eta_{\rm X}$ distribution may be more spread out, without
a clear bimodal distribution. This possibility is worth exploring
in the future.

Another mystery is regarding collimation of magnetar wind in short
GRBs. Our analysis suggests that at late times the magnetar wind
is essentially isotropic. On the other hand, during the prompt
emission phase, at least some short GRBs show evidence of collimation
(e.g. Burrows et al. 2006; Soderberg et al. 2006; Berger 2013 for a
review). There is no well studied short GRB prompt emission model
within the magnetar central engine scenario. Suggested scenarios invoke
an early brief accretion phase (Metzger et al. 2008), an early brief
differential rotation phase (Fan et al. 2013), or an early brief
phase-transition phase (e.g. Cheng \& Dai 1996; Chen \& Labun 2013).
The short GRB could be collimated by the torus within the accretion
scenario (Bucciantini et al. 2012).

Upon finishing this paper, we were drawn attention to Yi et al. (2014), who
performed an independent analysis on a sub-sample of GRB magnetar candidates
(essentially our Gold sample). They assumed that the long
GRB magnetar winds are isotropic and used the data to constrain magnetar
wind dissipation efficiencies. Through a
comparison with X-ray emission efficiency of spin-down powered pulsars,
they offer support to the millisecond magnetar central engine model
from a different point of view.

\acknowledgments

We acknowledge the use of the public data from the Swift data
archive and the UK Swift Science Data Center. We thank an anonymous
referee for helpful comments, He Gao and En-Wei Liang for helpful
discussion, and Zi-Gao Dai, Xue-Feng Wu, and Shuang-Xi Yi
for sharing their paper with us and related discussion. This work
is supported by the NASA ADAP program under grant NNX10AD48G.

\begin{center}
\begin{deluxetable}{lllllllllllll}
%\rotate
\tablewidth{480pt} \tabletypesize{\footnotesize}
\tabletypesize{\tiny} \tablecaption{The $\gamma$-ray and X-ray
observations and fitting results of the ``Gold'', ``Silver'',
and the short GRB samples.}\tablenum{1} \tablehead{
\colhead{GRB}& \colhead{$T_{90}$(s)}&
\colhead{$\Gamma_\gamma$\tablenotemark{a}}&
\colhead{$S_{\gamma,-7}$\tablenotemark{a}}&
\colhead{$\beta_{\rm x}$\tablenotemark{b}}& \colhead{($t_{1},
t_{2}$)(ks)\tablenotemark{c}}& \colhead{$t_{b}$
(ks)\tablenotemark{d}}& \colhead{$\alpha_1$\tablenotemark{e}}&
\colhead{$\alpha_2$\tablenotemark{e}}&\colhead{$\chi^2$/dof}&
\colhead{$z$\tablenotemark{f}}}

\startdata
Gold\\
\hline

060202	&198.9  &1.71$\pm$0.13  &21.3$\pm$1.65  &1.11$\pm$0.03 &(0.28,1.7)	&0.75$\pm$0.08 &0.23$\pm$0.03&5.79$\pm$0.16	
&563/521&---\\
060413	&147.7  &1.68$\pm$0.08  &35.6$\pm$1.47  &1.28$\pm$0.13  &(1.2,253.52)&26.43$\pm$1.12 &0.18$\pm$0.03	&3.42$\pm$0.21	
&79/71&---\\
060522	&71.1   &1.56$\pm$0.15  &11.4$\pm$1.11  &1.18$\pm$0.17  &(0.2,0.9)	&0.53$\pm$0.06  &0.14$\pm$0.36	&3.15$\pm$0.79	
&12/11&$5.11^{(1)}$\\
060607A	&102.2&1.47$\pm$0.08&25.5$\pm$1.12&0.67$\pm$0.06&(1.52,39.52)&12.34$\pm$0.19&0$\pm$0.01	&3.4$\pm$0.06	
&132/139&$3.082^{(1)}$\\
070110	&88.4   &1.58$\pm$0.12  &18$\pm$2	    &1.12$\pm$0.07  &(4.1,28.72) &20.4$\pm$0.44  &0.11$\pm$0.05	
&8.7$\pm$0.8&44/46&$2.352^{(1)}$\\
070616	&402.4  &1.61$\pm$0.04  &192$\pm$3.47   &0.26$\pm$0.01  &(0.13,2.01) &0.53$\pm$0.04  &-0.11$\pm$0.02 &5.29$\pm$0.05	
&224/241&---\\
090419	&450    &1.38$\pm$0.16  &25$\pm$2	    &0.30$\pm$0.28  &(0.12,1.72)	&0.49$\pm$0.07  &0.2$\pm$0.2	&3.44$\pm$0.23	 

&77/72&---\\
120213A	&49    &2.37$\pm$0.09  &19$\pm$1	    &0.95$\pm$0.21  &(1.04,12.84)&8.03$\pm$0.97  &0.35$\pm$0.06	
&4.56$\pm$0.24&49/53&---\\
130102A	&77.5   &1.39$\pm$0.18  &7.2$\pm$0.9    &0.80$\pm$0.41  &(0.18,10)	&0.42$\pm$0.26 &0.22$\pm$0.41	&5.92$\pm$0.57	
&12/10&---\\
\hline
Silver&(those&with&measured & redshifts & only)\\
\hline
050401	&33.3	&1.4$\pm$0.07	&82.2$\pm$3.06	&0.82$\pm$0.15	&(0.13,548)	&5.86$\pm$0.78	&0.57$\pm$0.02	&1.37$\pm$0.06	
&107/92&$2.9^{(1)}$\\
050505	&58.9	&1.41$\pm$0.12	&24.9$\pm$1.79	&1.23$\pm$0.04	&(2.88,133)	&7.87$\pm$1.57	&0.19$\pm$0.15	&1.3$\pm$0.06	
&27/45&$4.27^{(1)}$\\
050803	&87.9	&1.38$\pm$0.11	&21.5$\pm$1.35	&1.23$\pm$0.12	&(0.32,1330)&15.98$\pm$0.18&0.38$\pm$0.02	&1.89$\pm$0.06	
&95/75&$0.422^{(1)}$\\
060108	&14.3	&2.03$\pm$0.17	&3.69$\pm$0.37	&1.21$\pm$0.28	&(0.75,368)	&14.24$\pm$7.38	&0.12$\pm$0.08	&1.25$\pm$0.06	
&7/7&$2.03^{(1)}$\\
060526	&298.2	&2.01$\pm$0.24	&12.6$\pm$1.65	&1.16$\pm$0.16	&(1.02,314)	&10.02$\pm$4.55	&0.31$\pm$0.12	&1.5$\pm$0.23	
&34/48&$3.21^{(1)}$\\
060604	&95	    &2.01$\pm$0.42	&4.02$\pm$1.06	&1.15$\pm$0.17	&(1.23,824)	&11.37$\pm$6.8	&0.19$\pm$0.48	&1.17$\pm$0.08	
&35/41&$2.1357^{(1)}$\\
060605	&79.1	&1.55$\pm$0.2	&6.97$\pm$0.9	&1.36$\pm$0.12	&(0.15,103)	&7.45$\pm$0.52	&0.45$\pm$0.03	&2.01$\pm$0.05	
&16/21&$3.78^{(1)}$\\
060614	&108.7	&2.02$\pm$0.04	&204$\pm$3.63	&1.18$\pm$0.09	&(4.54,1795)&49.84$\pm$3.62	&0.18$\pm$0.06	&1.9$\pm$0.07	
&70/54&$0.125^{(1)}$\\
060729  &115.3	&1.75$\pm$0.14	&26.1$\pm$2.11	&1.24$\pm$0.03	&(0.52,8968)&72.97$\pm$3.02	&0.21$\pm$0.01	&1.42$\pm$0.02	
&160/459&$0.54^{(1)}$\\
060906	&43.5	&2.03$\pm$0.11	&22.1$\pm$1.36	&1.12$\pm$0.17	&(0.42,258)	&12.78$\pm$3.29&0.3$\pm$0.04	&1.81$\pm$0.1	
&5/7&$3.685^{(1)}$\\
060908	&19.3	&1.01$\pm$0.3	&28$\pm$1.11	&1.40$\pm$0.30	&(0.08,14.8)&0.71$\pm$0.17&0.43$\pm$0.09	&1.56$\pm$0.06	
&98/59&$1.8836^{(1)}$\\
061110A	&40.7	&1.67$\pm$0.12	&10.6$\pm$0.76	&1.10$\pm$0.32	&(3.08,756)	&73.17$\pm$5.67&0.19$\pm$0.15	&1.16$\pm$0.17	
&7/5&$0.758^{(1)}$\\
070306	&209.5	&1.66$\pm$0.1	&53.8$\pm$2.86	&1.19$\pm$0.08	&(0.48,819)	&29.69$\pm$1.72	&0.12$\pm$0.02	&1.87$\pm$0.03	
&153/132&$1.497^{(1)}$\\
070529	&109.2	&1.34$\pm$0.16	&25.7$\pm$2.45	&0.76$\pm$0.24	&(0.17,445)	&1.65$\pm$0.84	&0.64$\pm$0.07	&1.36$\pm$0.05	
&23/19&$2.4996^{(1)}$\\
080605	&20	    &1.11$\pm$0.14	&133$\pm$2	    &0.74$\pm$0.16	&(0.09,101)	&0.44$\pm$0.05&0.5$\pm$0.05	&1.34$\pm$0.02	
&330/289&$1.6398^{(1)}$\\
080607	&79	    &1.31$\pm$0.04	&240$\pm$9	    &1.13$\pm$0.15	&(0.62,401)	&1.38$\pm$0.19&0.05$\pm$0.33	&1.68$\pm$0.04	
&103/98&$3.036^{(1)}$\\
080721	&16.2	&1.11$\pm$0.08	&120$\pm$10 	&0.84$\pm$0.06	&(0.11,2011)&3.09$\pm$0.16	&0.8$\pm$0.01	&1.65$\pm$0.02	
&54/49&$2.602^{(1)}$\\
080905B	&128	&1.78$\pm$0.15	&18$\pm$2   	&1.22$\pm$0.10	&(0.22,988)	&4.03$\pm$1.22	&0.25$\pm$0.03	&1.46$\pm$0.02	
&94/98&$2.374^{(2)}$\\
081008	&185.5	&1.69$\pm$0.07	&43$\pm$2	    &0.98$\pm$0.11	&(0.71,502)	&15.92$\pm$6.58&0.81$\pm$0.03	&1.85$\pm$0.08	
&33/38&$1.9685^{(3)}$\\
081203A	&294	&1.54$\pm$0.06	&77$\pm$3    	&1.04$\pm$0.11	&(0.2,506)	&11.23$\pm$8.69&1.12$\pm$0.01	&2.07$\pm$0.07	
&191/163&$2.1^{(4)}$\\
081221	&34 	&1.21$\pm$0.13	&181$\pm$3	    &1.29$\pm$0.10	&(0.25,498)	&0.6$\pm$0.08	&0.3$\pm$0.11	&1.32$\pm$0.02	
&285/312&$2.26^{(5)}$\\
090423	&10.3	&0.8$\pm$0.5	&5.9$\pm$0.4	&0.92$\pm$0.16	&(0.39,501)	&4.28$\pm$0.76&-0.16$\pm$0.07	&1.42$\pm$0.04	
&27/33&$8.2^{(6)}$\\
090618	&113.2	&1.42$\pm$0.09	&1050$\pm$10	&0.72$\pm$0.05	&(0.58,1998)&7.28$\pm$1.43	&0.67$\pm$0.02	&1.48$\pm$0.03	
&128/132&$0.54^{(7)}$\\
090927	&2.2	&1.8$\pm$0.2	&2$\pm$	0.3     &0.92$\pm$0.23 	&(2.52,1003)&8.29$\pm$1.32	&0.16$\pm$0.11	&1.24$\pm$0.09	
&19/15&$1.37^{(8)}$\\
091208B	&14.9	&1.74$\pm$0.11	&33$\pm$2   	&1.04$\pm$0.16 	&(0.14,969)	&1.15$\pm$0.21	&0.16$\pm$0.14	&1.17$\pm$0.03	
&79/68&$1.063^{(9)}$\\
100418A	&7	    &2.16$\pm$0.25	&3.4$\pm$0.5	&1.27$\pm$0.23	&(0.51,2002)&86.82$\pm$22.14&-0.11$\pm$0.05&1.53$\pm$0.06	
&44/49&$0.6235^{(10)}$\\
111008A	&63.5	&1.86$\pm$0.09	&53$\pm$3   	&1.07$\pm$0.23	&(0.31,987)	&7.47$\pm$2.28	&0.29$\pm$0.02	&1.34$\pm$0.02	
&143/167&$4.9898^{(11)}$\\
111228A	&101.2	&2.27$\pm$0.06	&85$\pm$2   	&1.12$\pm$0.08	&(0.42,2990)    &6.53$\pm$2.11	&0.22$\pm$0.03	&1.23$\pm$0.01	 

&202/187&$0.7156^{(12)}$\\
120422A	&5.35	&1.19$\pm$0.24	&2.3$\pm$0.4	&1.22$\pm$0.23	&(0.49,2011)&166.15$\pm$22.33&0.27$\pm$0.04&1.27$\pm$0.14	
&4/6&$0.283^{(13)}$\\
121024A	&69	    &1.41$\pm$0.22	&11$\pm$1	    &0.94$\pm$0.14	&(2.01,504)	&32.98$\pm$8.21&0.8$\pm$0.06	&1.71$\pm$0.09	
&47/52&$2.298^{(14)}$\\
121027A	&62.6	&1.82$\pm$0.09	&20$\pm$1	    &1.45$\pm$0.11	&(40.1,3019)&144.71$\pm$44.87&0.37$\pm$0.07&1.52$\pm$0.05	
&54/46&$1.773^{(15)}$\\
121128A	&23.3	&1.32$\pm$0.18	&69$\pm$4	    &1.32$\pm$0.21	&(0.21,98.7)&1.58$\pm$0.24&0.52$\pm$0.07	&1.68$\pm$0.04	
&81/78&$2.2^{(16)}$\\
121229A	&100	&2.43$\pm$0.46	&4.6$\pm$1.3	&1.10$\pm$0.30	&(2.04,205)	&56.39$\pm$8.34&0.21$\pm$0.12	&1.43$\pm$0.27	
&3/5&$2.707^{(17)}$\\
\hline
SGRBs&(those&with&measured & redshifts & only)\\
\hline
051221A	&1.4	&1.39$\pm$0.06	&11.5$\pm$0.35	&1.07$\pm$0.13	&(6.02,655) &34.32$\pm$6.78 &0.19$\pm$0.08 &1.45$\pm$0.05	
&41/44&$0.55^{(18)}$\\
060801	&0.49	&1.27$\pm$0.16	&0.8$\pm$0.1   	&0.43$\pm$0.12	&(0.08,0.73)&0.06$\pm$0.04	&0.67$\pm$0.12 &4.81$\pm$0.62	
&22/18&$1.131^{(18)}$\\
061201	&0.6	&0.81$\pm$0.15	&3.24$\pm$0.27  &1.2$\pm$0.22	&(0.11,30.9)&1.21$\pm$0.26	&0.52$\pm$0.06 &1.87$\pm$0.07	
&16/18&$0.111^{(18)}$\\
070809	&1.3	&1.69$\pm$0.22	&1.0$\pm$0.1	&0.37$\pm$0.21	&(0.53,67.4)&12.86$\pm$6.52 &-0.01$\pm$0.09&1.14$\pm$0.13	
&33/26&$0.219^{(18)}$\\
090426	&1.2	&1.93$\pm$0.22	&1.8$\pm$0.3	&1.04$\pm$0.15	&(0.13,17.6)&0.31$\pm$0.18  &-0.18$\pm$0.16&1.02$\pm$0.04	
&25/19&$2.6^{(18)}$\\
090510	&0.3	&0.98$\pm$0.21	&3.4$\pm$0.4	&0.75$\pm$0.12	&(0.11,20.7)&0.28$\pm$0.04  &0.62$\pm$0.03 &2.17$\pm$0.05	
&76/68&$0.903^{(18)}$\\
100724A	&1.4	&1.92$\pm$0.21	&1.6$\pm$0.2	&0.94$\pm$0.23	&(0.38,0.89)&0.52$\pm$0.16  &0.21$\pm$0.12 &1.84$\pm$0.51	
&45/33&$1.288^{(19)}$\\
101219A	&0.6	&0.63$\pm$0.09	&4.6$\pm$0.3	&0.53$\pm$0.26	&(0.05,0.27)&0.23$\pm$0.15  &0.21$\pm$0.24 &6.82$\pm$0.96	
&38/29&$0.718^{(18)}$\\
130603B	&0.18	&1.83$\pm$0.12	&19.2$\pm$1.2	&1.18$\pm$0.18	&(0.07,48.1)&3.01$\pm$0.67  &0.38$\pm$0.02 &1.64$\pm$0.04	
&111/98&$0.356^{(20)}$\\

\enddata

\tablenotetext{a}{The photon index and
gamma-ray fluence in the BAT band (15-150keV, in units of
$10^{-7}$ erg cm$^{-2}$).} \tablenotetext{b}{The spectral index
of the absorbed power-law model for the plateau or the normal segments.}
\tablenotetext{c}{Time interval (from $t_1$ to $t_2$) of our XRT light curve
fitting; times in units of kilo seconds.} \tablenotetext{d}{The break
time of the lightcurves from our
fitting.}\tablenotetext{e}{$\alpha_1$ and $\alpha_2$ are the
decay slopes before and after the break time.}
\tablenotetext{f}{The References of redshift measurements.}

\tablerefs{1: Evans et al.(2009); 2: Vreeswijk et al.(2008); 3:
D'Avanzo et al.(2008); 4: Landsman et al.(2008); 5: Salvaterra
et al.(2012); 6: Tanvir et al.(2009); 7: Cenko et al.(2009); 8:
Levan et al.(2009); 9: Wiersema et al.(2009); 10: Antonelli et
al.(2010); 11: Wiersema et al.(2011); 12: Cucchiara et
al.(2011); 13: Schulze et al.(2012); 14: Tanvir et al.(2012);
15: Levan et al.(2012); 16: Tanvir et al.(2012); 17: Fynbo et
al.(2012); 18: Rowlinson et al.(2013); 19: Thoene et al.(2010);
20: Fong et al.(2014).}
\end{deluxetable}
\end{center}

\begin{center}
\begin{deluxetable}{lllllllllllll}
%\rotate
\tablewidth{500pt} \tabletypesize{\footnotesize}
\tabletypesize{\tiny} \tablecaption{The properties of GRBs with
known redshifts in our ``Gold'', ``Silver'', and short GRB
samples.}\tablenum{2}

\tablehead{ \colhead{GRB}&
\colhead{$\theta_{j}$\tablenotemark{a}}& \colhead{$E_{\rm
\gamma,iso,52}$\tablenotemark{b}}&
\colhead{$L_{b,49}$\tablenotemark{c}}&
\colhead{$\tau_{3}$\tablenotemark{c}}&
\colhead{$B_{p,15}$\tablenotemark{d}}&
\colhead{$P_{0,-3}$\tablenotemark{d}}&
\colhead{$B_{p,\theta,15}$\tablenotemark{e}}&
\colhead{$P_{0,\theta,-3}$\tablenotemark{e}}& \colhead{$E_{\rm
rot,50}$\tablenotemark{f}}}

\startdata
Gold\\
\hline
060522	&5	&0.71$\pm$0.71 &1.38$\pm$0.22 &0.06$\pm$0.01 	&2.34$\pm$0.71 &1.19$\pm$0.59
&37.93$\pm$11.42&19.28$\pm$9.51&0.54$\pm$0.29\\
060607A	&5	&9.08$\pm$7.11 &0.58$\pm$0.07 &0.13$\pm$0.02 	&0.18$\pm$0.04 &0.44$\pm$0.10 &2.91$\pm$0.67  &7.15$\pm$1.58
&3.91$\pm$1.29\\
070110	&5	&3.09$\pm$2.51 &0.07$\pm$0.03 &3.68$\pm$0.06 	&0.23$\pm$0.11 &0.74$\pm$0.36 &3.79$\pm$1.78
&11.97$\pm$5.84&1.39$\pm$0.76\\
\hline
Silver\\
\hline
050401	&5	&$32^{+26}_{-7}$&0.47$\pm$0.01 &1.51$\pm$0.21 &0.09$\pm$0.02 	&0.15$\pm$0.02 	&1.44$\pm$0.32&2.43$\pm$0.34
&34.20$\pm$8.11\\
050505	&$1.67\pm0.35$	&$16^{+13}_{-3}$ &0.41$\pm$0.01 &1.49$\pm$0.30 &0.07$\pm$0.02 	&0.13$\pm$0.03 	&3.21$\pm$1.20 	
&6.28$\pm$1.45&5.12$\pm$1.78\\
050803	&5	&$0.24^{+0.24}_{-0.08}$ &(8.92$\pm$0.31)e-4 &11.24$\pm$0.13 &0.07$\pm$0.01 	&0.21$\pm$0.01&1.22$\pm$0.04
&3.39$\pm$0.08&17.37$\pm$7.88\\
060108	&5	&$0.59^{+0.84}_{-0.08}$ &(5.96$\pm$0.48)e-3 &4.70$\pm$2.44 &0.21$\pm$0.36 	&0.53$\pm$0.50 	
&3.28$\pm$0.58&8.65$\pm$2.43&2.68$\pm$1.79\\
060526	&$3.61\pm0.57$	&$5.2^{+5.6}_{-0.4}$&(3.91$\pm$0.27)e-2 &2.38$\pm$1.08 &0.10$\pm$0.09 	&0.21$\pm$0.18 	
&2.14$\pm$1.59&4.74$\pm$1.98&8.93$\pm$2.69\\
060604	&5	&$0.5^{+0.12}_{-0.1}$ &(1.15$\pm$0.05)e-2 &3.63$\pm$2.17& 0.21$\pm$0.08 	&0.49$\pm$0.36 	
&3.39$\pm$1.87&7.98$\pm$3.88&3.16$\pm$2.45\\
060605	&$1.55\pm0.57$	&$2.5^{+3.1}_{-0.6}$ &0.13$\pm$0.01 &1.56$\pm$0.11 &0.03$\pm$0.01 	&0.06$\pm$0.01 	
&1.68$\pm$0.23&3.19$\pm$0.31&19.57$\pm$3.35\\
060614	&$7.57\pm2.29$	&$0.24^{+0.04}_{-0.04}$ &(2.49$\pm$0.08)e-5 &44.31$\pm$3.22 &0.06$\pm$0.01 	
&0.32$\pm$0.03&0.69$\pm$0.10&3.42$\pm$0.34&17.06$\pm$2.92\\
060729	&$18\pm1.61$	&$0.33^{+0.29}_{-0.06}$&(1.56$\pm$0.02)e-3 &47.38$\pm$1.96 &0.06$\pm$0.01 	
&0.33$\pm$0.02&0.25$\pm$0.02&1.48$\pm$0.07&91.23$\pm$7.84\\
060906	&$1.15\pm0.12$	&$13^{+12}_{-1}$ &(3.08$\pm$0.21)e-2 &2.73$\pm$0.70 &0.04$\pm$0.02 	&0.09$\pm$0.04 	
&2.55$\pm$1.70&6.36$\pm$2.78&4.94$\pm$2.55\\
060908	&$0.46\pm0.29$	&$7^{+4}_{-1}$ &0.26$\pm$0.07 &0.25$\pm$0.06 &0.57$\pm$0.41 	&0.33$\pm$0.17 	
&100.6$\pm$73.24&59.08$\pm$30.06&0.06$\pm$0.03\\
061110A	&5	&$0.28^{+0.28}_{-0.06}$ &(3.83$\pm$0.81)e-5 &41.62$\pm$3.23 &0.39$\pm$0.09 	
&2.32$\pm$0.41&6.31$\pm$1.43&37.68$\pm$6.71&0.14$\pm$0.04\\
070306	&$3.38\pm1.72$	&$6^{+5}_{-1}$ &(2.06$\pm$0.06)e-2 &11.89$\pm$0.69 &0.02$\pm$0.01 	&0.06$\pm$0.01 	
&0.36$\pm$0.04&1.39$\pm$0.11&104.2$\pm$14.5\\
070529	&5	&$9^{+9}_{-3}$ &0.12$\pm$0.01 &0.47$\pm$0.24 &0.44$\pm$0.17 	&0.39$\pm$0.33 	&7.06$\pm$1.58&6.33$\pm$3.63
&5.02$\pm$3.55\\
080605	&5	&$21^{+9}_{-4}$ &1.49$\pm$0.16 &0.17$\pm$0.02 &0.47$\pm$0.11 	&0.22$\pm$0.03 	&7.63$\pm$1.75&3.52$\pm$0.56
&16.26$\pm$4.22\\
080607	&5	&$280^{+130}_{-90}$&1.36$\pm$0.27 &0.34$\pm$0.05 &0.13$\pm$0.05 	&0.11$\pm$0.03 	&2.10$\pm$0.77&1.72$\pm$0.47
&67.85$\pm$26.14\\
080721	&5	&$110^{+110}_{-50}$&2.50$\pm$0.13 &0.86$\pm$0.04 &0.07$\pm$0.01 	&0.09$\pm$0.01 	&1.18$\pm$0.12 	
&1.45$\pm$0.10&96.86$\pm$13.41\\
080905B	&5	&$3.4^{+3.1}_{-0.6}$ &0.31$\pm$0.01 &1.20$\pm$0.36 &0.11$\pm$0.08 	&0.16$\pm$0.07 	&1.84$\pm$1.30&2.58$\pm$1.10
&30.32$\pm$15.55\\
081008	&5	&$6^{+3}_{-1}$ &(1.19$\pm$0.06)e-2 &5.36$\pm$2.22 &0.08$\pm$0.10 	&0.22$\pm$0.16 	&1.29$\pm$0.63&3.60$\pm$1.63
&15.45$\pm$10.29\\
081203A	&5	&$17^{+13}_{-4}$ &(3.23$\pm$0.11)e-2 &3.62$\pm$2.80 &0.06$\pm$0.01 	&0.13$\pm$0.07 	
&0.93$\pm$0.39&2.17$\pm$0.73&42.65$\pm$4.11\\
081221	&5	&282.29$\pm$4.68&1.72$\pm$0.20 &0.18$\pm$0.02 &0.38$\pm$0.11 	&0.21$\pm$0.04 	&6.19$\pm$1.79&3.35$\pm$0.67
&18.01$\pm$5.61\\
090423	&$>12$	&$8^{+1}_{-1}$ &0.82$\pm$0.04 &0.47$\pm$0.08 &0.19$\pm$0.07 	&0.28$\pm$0.07 	&1.30$\pm$0.47&1.88$\pm$0.44
&59.46$\pm$22.42\\
090618	&$6.7\mp1.08$	&$15^{+1}_{-1}$ &(1.62$\pm$0.02)e-2 &4.73$\pm$0.93 &0.25$\pm$0.09 	&0.48$\pm$0.13 	
&3.08$\pm$1.09&5.79$\pm$1.23&5.98$\pm$1.94\\
090927	&5	&0.43$\pm$0.06 	&(5.16$\pm$0.27)e-3 &55.36$\pm$32.68 &0.05$\pm$0.02 	&0.36$\pm$0.25 	 &0.73$\pm$0.21&5.82$\pm$0.86
&6.01$\pm$1.04\\
091208B	&$7.3\pm1.42$	&4.88$\pm$0.30 	&0.05$\pm$0.01 &0.56$\pm$0.10 &0.96$\pm$0.60 &0.72$\pm$0.34 	
&10.71$\pm$6.65&8.02$\pm$3.77&3.11$\pm$1.67\\
100418A	&5	&0.14$\pm$0.02 	&(1.16$\pm$0.11)e-4 &53.48$\pm$13.64 &0.05$\pm$0.03 	&0.32$\pm$0.13&0.80$\pm$0.51&5.23$\pm$2.16
&7.31$\pm$3.65\\
111008A	&5	&85.23$\pm$4.82 &0.72$\pm$0.02 &1.25$\pm$0.38 &0.05$\pm$0.04 	&0.11$\pm$0.04 	&0.88$\pm$0.59 	&1.67$\pm$0.67
&72.35$\pm$35.86\\
111228A	&5	&5.45$\pm$0.13 	&(8.48$\pm$0.24)e-3 &3.81$\pm$1.23 &0.36$\pm$0.25 	&0.64$\pm$0.26 	&5.78$\pm$1.19&10.32$\pm$4.22
&1.89$\pm$0.94\\
120422A	&5	&0.13$\pm$0.02 	&(2.31$\pm$0.26)e-6 &129.5$\pm$17.4 &0.42$\pm$0.12 	&3.80$\pm$0.75&6.85$\pm$1.96&61.66$\pm$12.09
&0.05$\pm$0.02\\
121024A	&5	&10.78$\pm$0.98 &(6.30$\pm$0.45)e-3 &10.01$\pm$2.49 &0.06$\pm$0.02 	&0.26$\pm$0.09 	&1.05$\pm$0.58&4.19$\pm$1.48
&11.38$\pm$5.16\\
121027A	&5	&6.61$\pm$0.33 	&(3.38$\pm$0.16)e-3 &52.19$\pm$16.18 &0.02$\pm$0.01 	&0.15$\pm$0.07 	 &0.29$\pm$0.15&2.41$\pm$1.09
&34.58$\pm$18.26\\
121128A	&5	&78.91$\pm$4.57 &0.38$\pm$0.06 &0.50$\pm$0.08 &0.21$\pm$0.08 	&0.18$\pm$0.05 	&3.39$\pm$1.29&2.98$\pm$0.82
&22.61$\pm$8.72\\
121229A	&5	&6.64$\pm$1.88 	&(1.81$\pm$0.25)e-3 &15.21$\pm$2.25 &0.07$\pm$0.02 	&0.34$\pm$0.08 	&1.06$\pm$0.36&5.55$\pm$1.32
&6.52$\pm$2.27\\
\hline
SGRBs\\
\hline
051221A	&-  &$0.28^{+0.21}_{-0.11}$  	&24.71$\pm$4.97&(8.8$\pm$0.23)e-3 &0.57$\pm$0.01 	&2.47$\pm$0.16 	&-&-
&22.88$\pm$4.41\\
060801	&-  &$0.17^{+0.02}_{-0.02}$ 	&0.03$\pm$0.02 &8.7$\pm$4.1       &11.21$\pm$4.21 	&1.95$\pm$0.34  &-&-
&52.62$\pm$24.26\\
061201	&-  &$0.018^{+0.002}_{-0.001}$  &1.08$\pm$0.23 &0.08$\pm$0.01     &6.01$\pm$0.12 	&4.59$\pm$0.05 	&-&-
&9.48$\pm$1.88\\
070809	&-	&$0.001^{+0.001}_{-0.001}$ 	&12.14$\pm$5.33 &(4.5$\pm$2.5)e-3  &2.06$\pm$1.03 	&5.55$\pm$1.25 	&-&-
&6.49$\pm$4.31\\
090426	&-	&$0.42^{+0.5}_{-0.04}$ 	    &0.09$\pm$0.05 &1.9$\pm$1.2       &4.79$\pm$3.11 	&1.87$\pm$0.53  &-&-
&57.46$\pm$5.44\\
090510	&-	&$0.3^{+0.5}_{-0.2}$        &0.15$\pm$0.02 &2.1$\pm$0.2       &5.05$\pm$0.26 	&1.87$\pm$0.05 	&-&-
&57.36$\pm$3.02\\
100724A	&-	&$0.07^{+0.01}_{-0.01}$ 	&0.23$\pm$0.07 &0.23$\pm$0.03     &8.22$\pm$0.59 	&4.14$\pm$0.15 	&-&-
&11.67$\pm$1.87\\
101219A	&-	&$0.48^{+0.03}_{-0.03}$     &0.13$\pm$0.06 &9.7$\pm$3.8       &2.86$\pm$0.81 	&0.96$\pm$0.13 	&-&-
&217.7$\pm$71.5\\
130603B	&-	&$0.22^{+0.02}_{-0.02}$ 	&2.22$\pm$0.49 &0.11$\pm$0.01     &2.16$\pm$0.12 	&2.61$\pm$0.07 	&-&-
&29.32$\pm$1.63\\
\enddata

\tablenotetext{a}{The jet opening angle (in units of degree
$(^{\circ})$) measured from afterglow observations  (Racusin et
al. 2009; Lu et al. 2011; Nemmen et al. 2012), or assumed as
$\theta_{j}=5^{\circ}$ if no observation is available. SGRBs
are assumed to be isotropic.} \tablenotetext{b}{$E_{\rm \gamma,
iso}$ is calculated using fluence and redshift extrapolated
into 1-10000 keV (rest frame) with a spectral model and a
k-correction, in units of $10^{52}$ erg.}
\tablenotetext{c}{Isotropic luminosity of break time (in units
of $10^{49}~{\rm erg~s^{-1}}$), and the spin-down time (in
units of $10^3$s).} \tablenotetext{d}{Dipolar magnetic field
strength at the polar cap in units of $10^{15} G$, and the
initial spin period of the magnetar in units of milliseconds,
with an assumption of an isotropic wind.} \tablenotetext{e}{The
same as $d$, but with beaming correction made.}
\tablenotetext{f}{The rotational energy (in units of $10^{50}$
erg) of the magnetar assuming $R_{6}=1$ and $M=1.4M_{\odot}$.}
\end{deluxetable}
\end{center}

\begin{center}
\begin{deluxetable}{lllllllllllll}
%\rotate
\tablewidth{240pt} \tabletypesize{\footnotesize}
%\tabletypesize{\tiny}
\tablecaption{The center value of Gaussian fitting of the distributions.}\tablenum{3}

\tablehead{ \colhead{}& \colhead{Gold+Silver}&
\colhead{Gold+Silver+Aluminum}& \colhead{Non-magnetar} }

\startdata
$E_{\rm \gamma,iso}$ &($52.87\pm0.33$) erg    &($52.89\pm0.09$) erg     &($53.20\pm0.04$) erg\\
$E_{\rm K,iso}$      &($53.11\pm0.09$) erg    &($53.99\pm0.06$) erg     &($53.94\pm0.02$) erg\\
$E_{\rm total,iso}$  &($53.31\pm0.05$) erg    &($54.05\pm0.05$) erg     &($54.01\pm0.05$) erg\\
\hline
&Silver & Silver+Aluminum & Non-magnetar\\
\hline
$E_{\gamma}$     &($48.55\pm0.11$) erg    &($49.06\pm0.13$) erg     &($50.11\pm0.12$) erg\\
$E_{\rm K}$      &($50.55\pm0.17$) erg    &($51.13\pm0.12$) erg     &($51.54\pm0.18$) erg\\
$E_{\rm total}$  &($50.62\pm0.07$) erg    &($51.06\pm0.09$) erg     &($51.81\pm0.11$) erg\\
\enddata

\end{deluxetable}
\end{center}

\begin{figure}
\includegraphics[angle=0,scale=0.35]{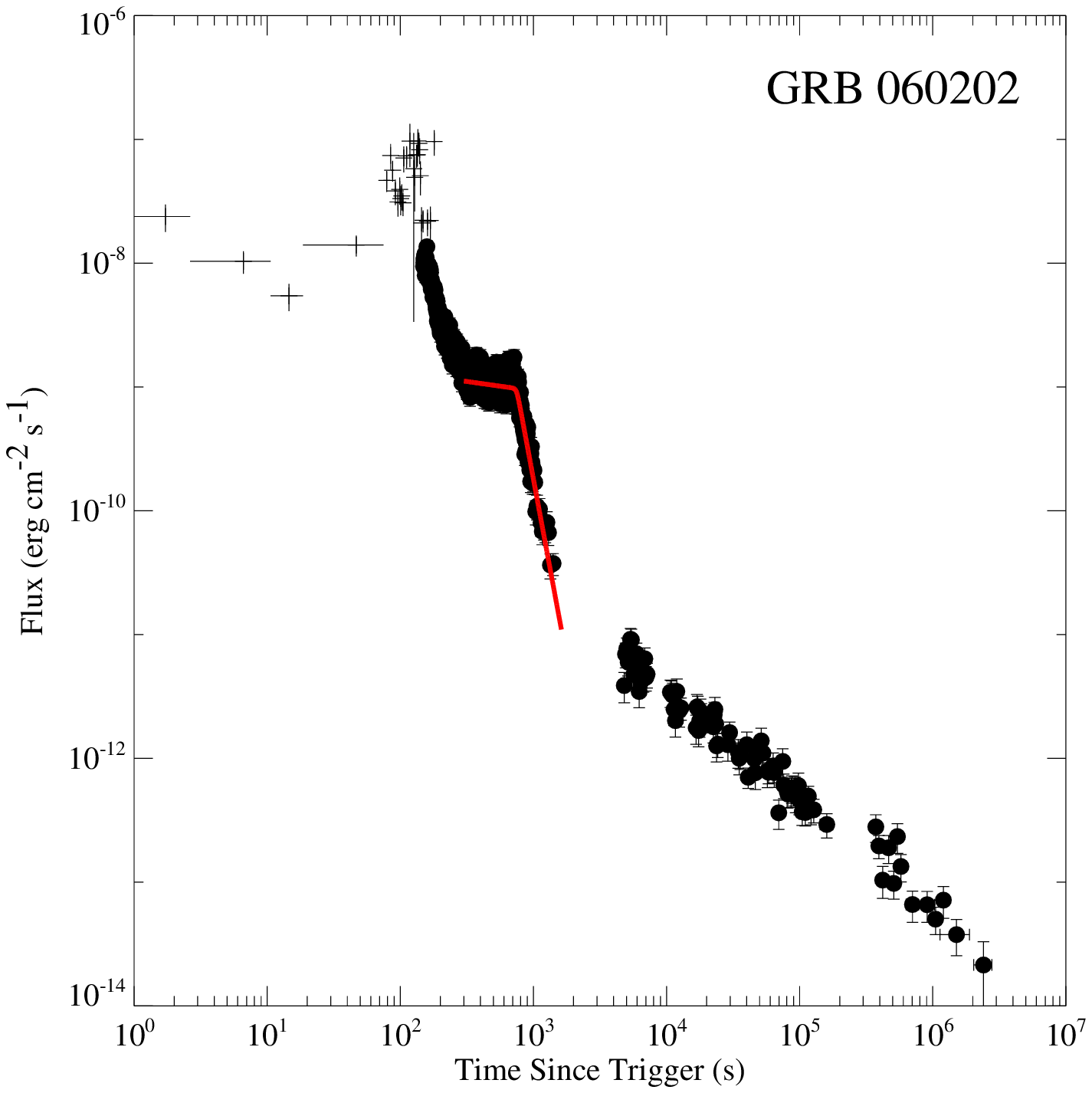}
\includegraphics[angle=0,scale=0.35]{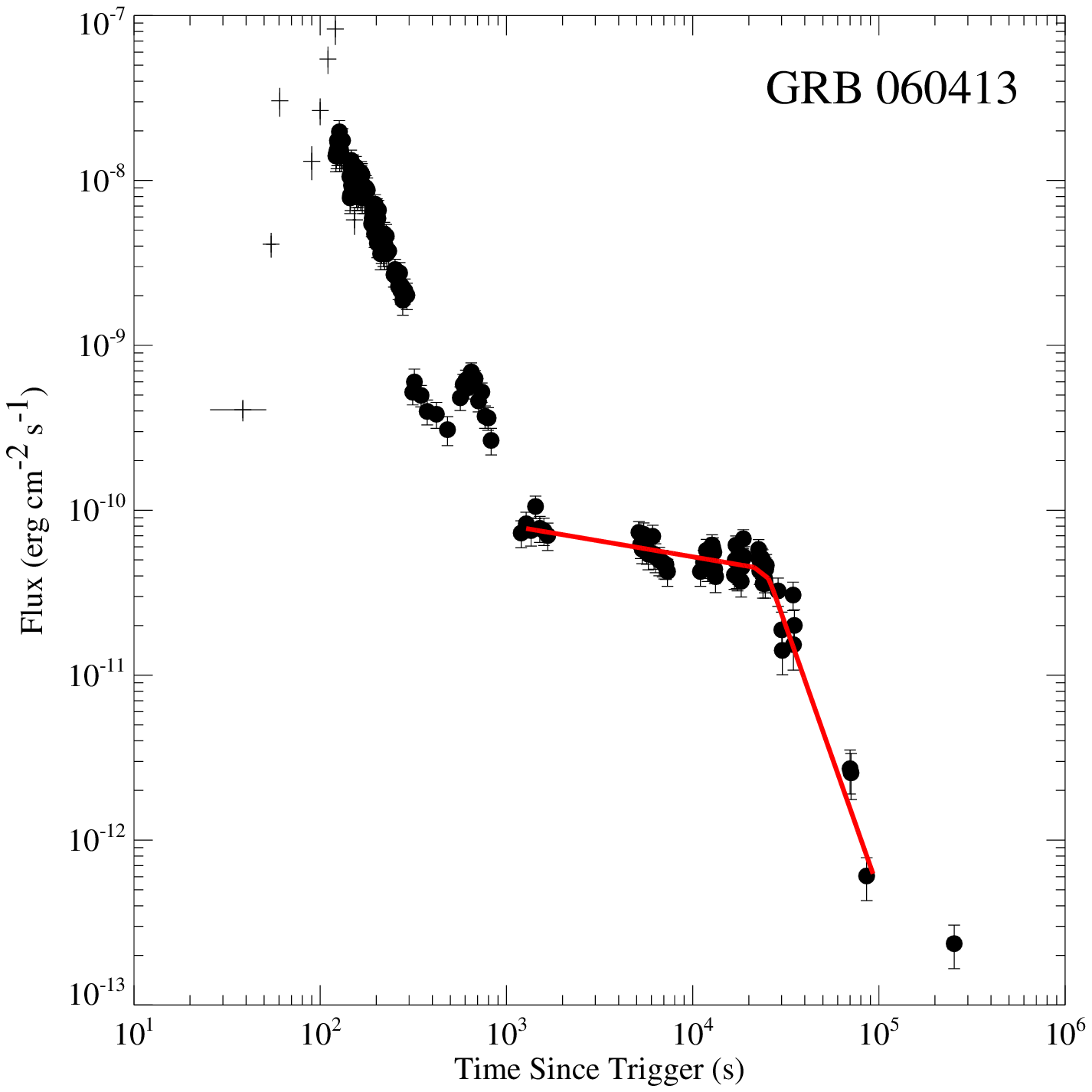}
\includegraphics[angle=0,scale=0.35]{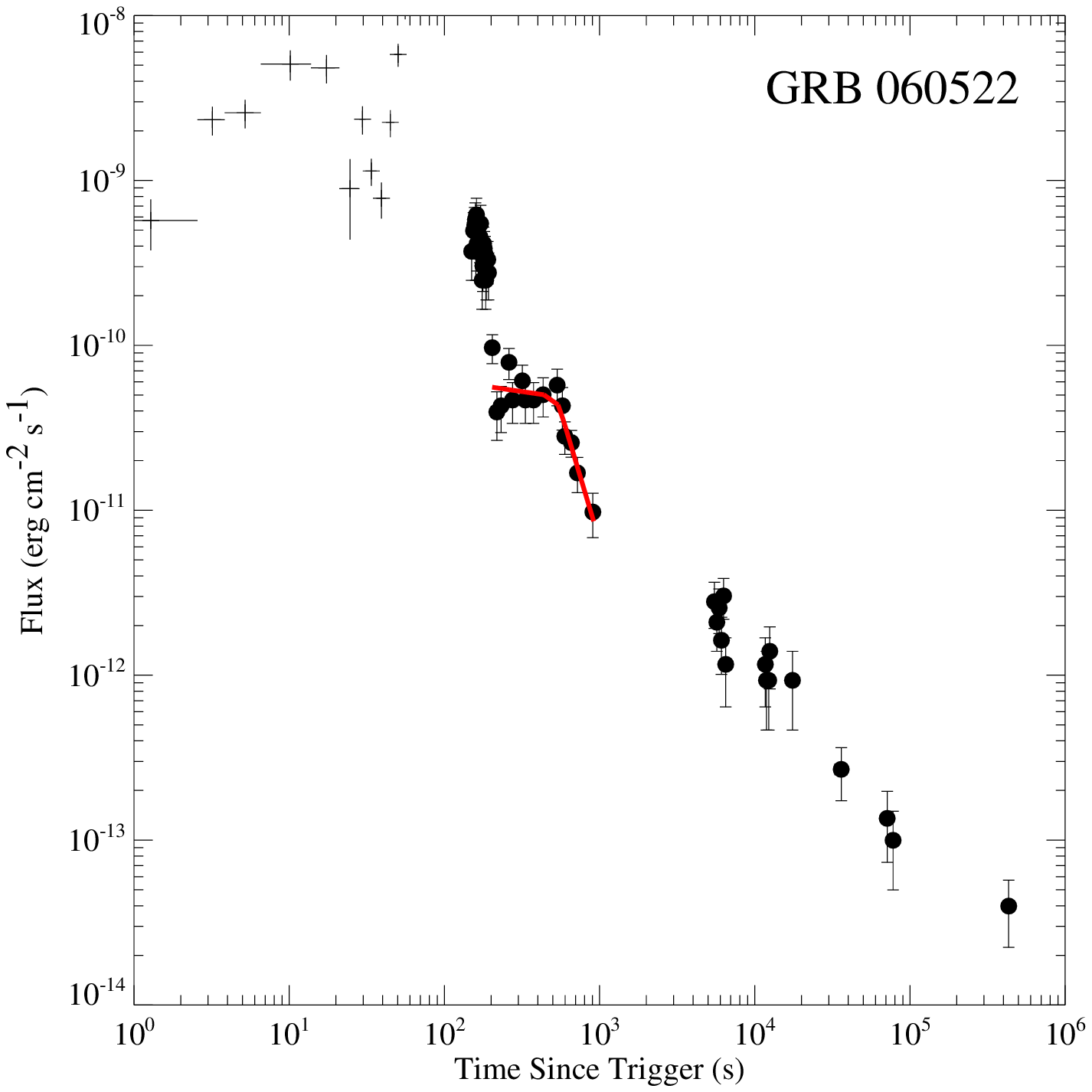}
\includegraphics[angle=0,scale=0.35]{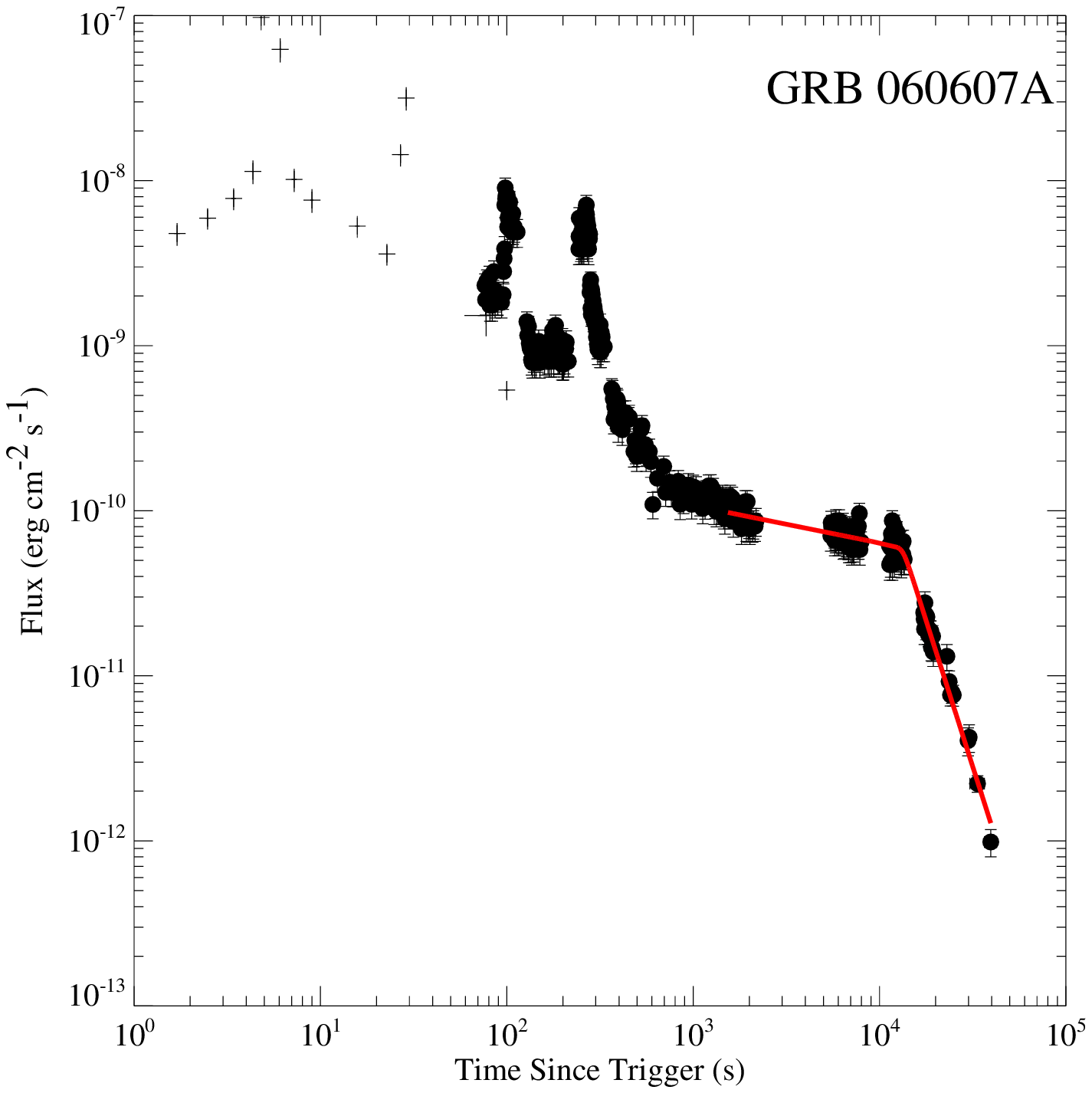}
\includegraphics[angle=0,scale=0.35]{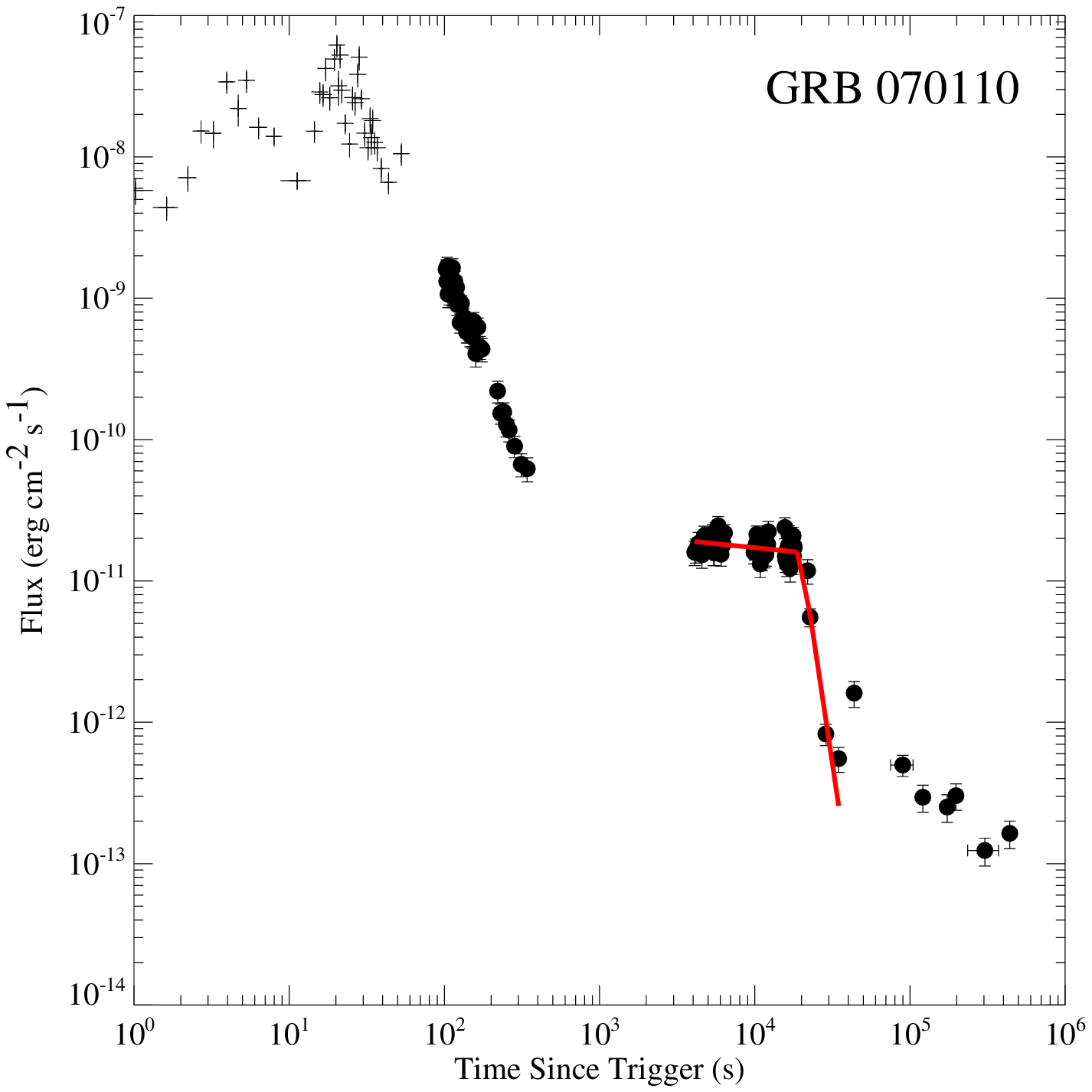}
\includegraphics[angle=0,scale=0.35]{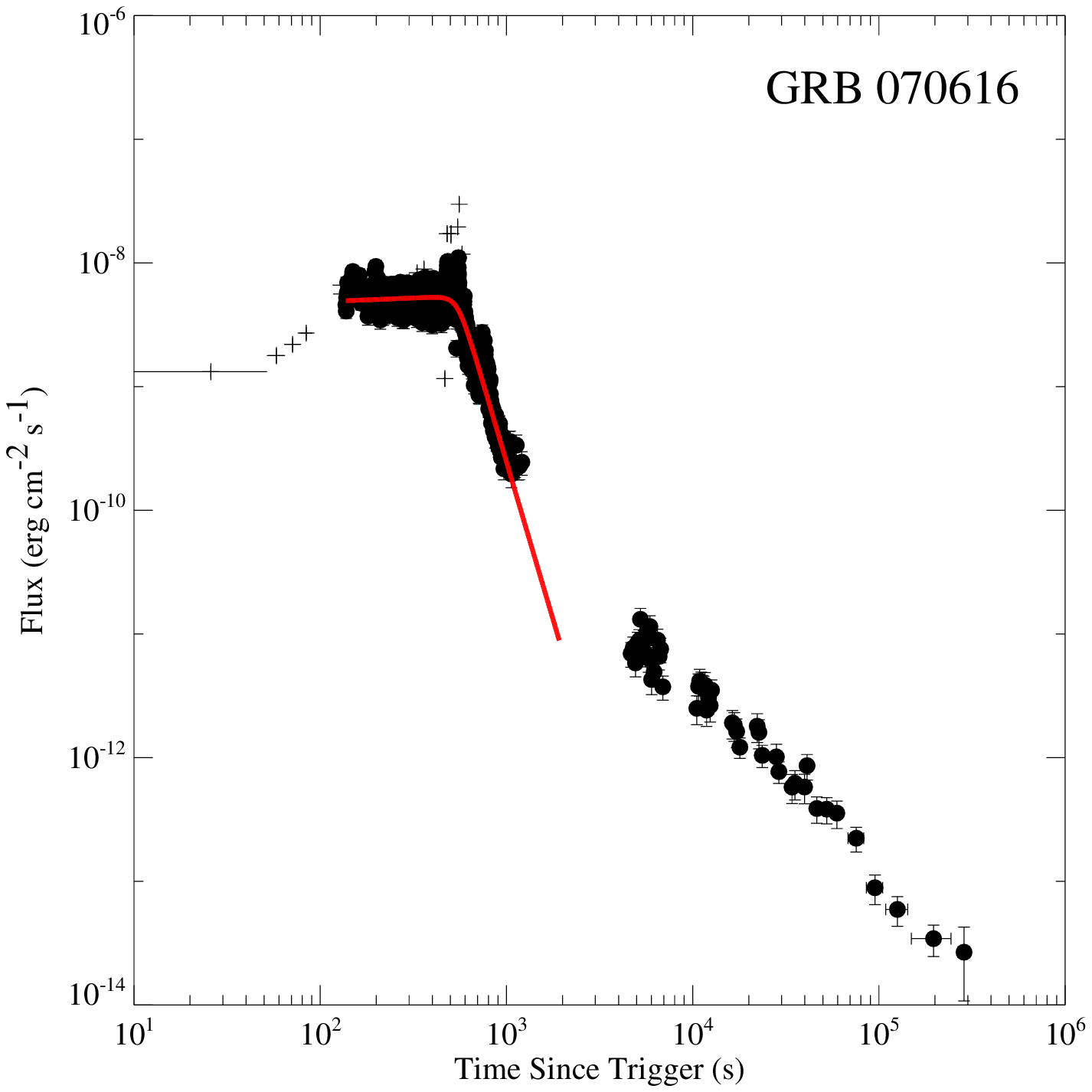}
\includegraphics[angle=0,scale=0.35]{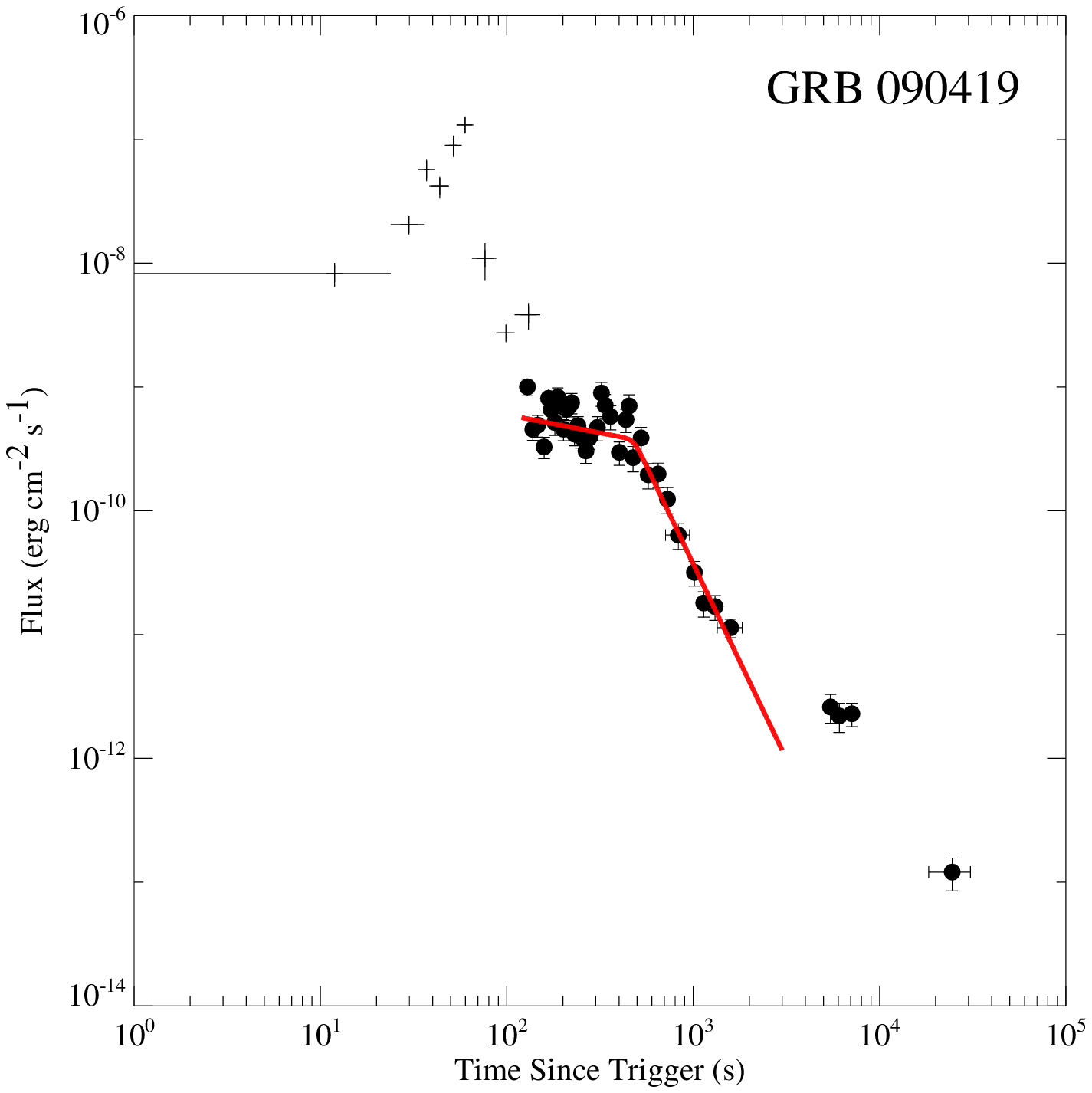}
\includegraphics[angle=0,scale=0.35]{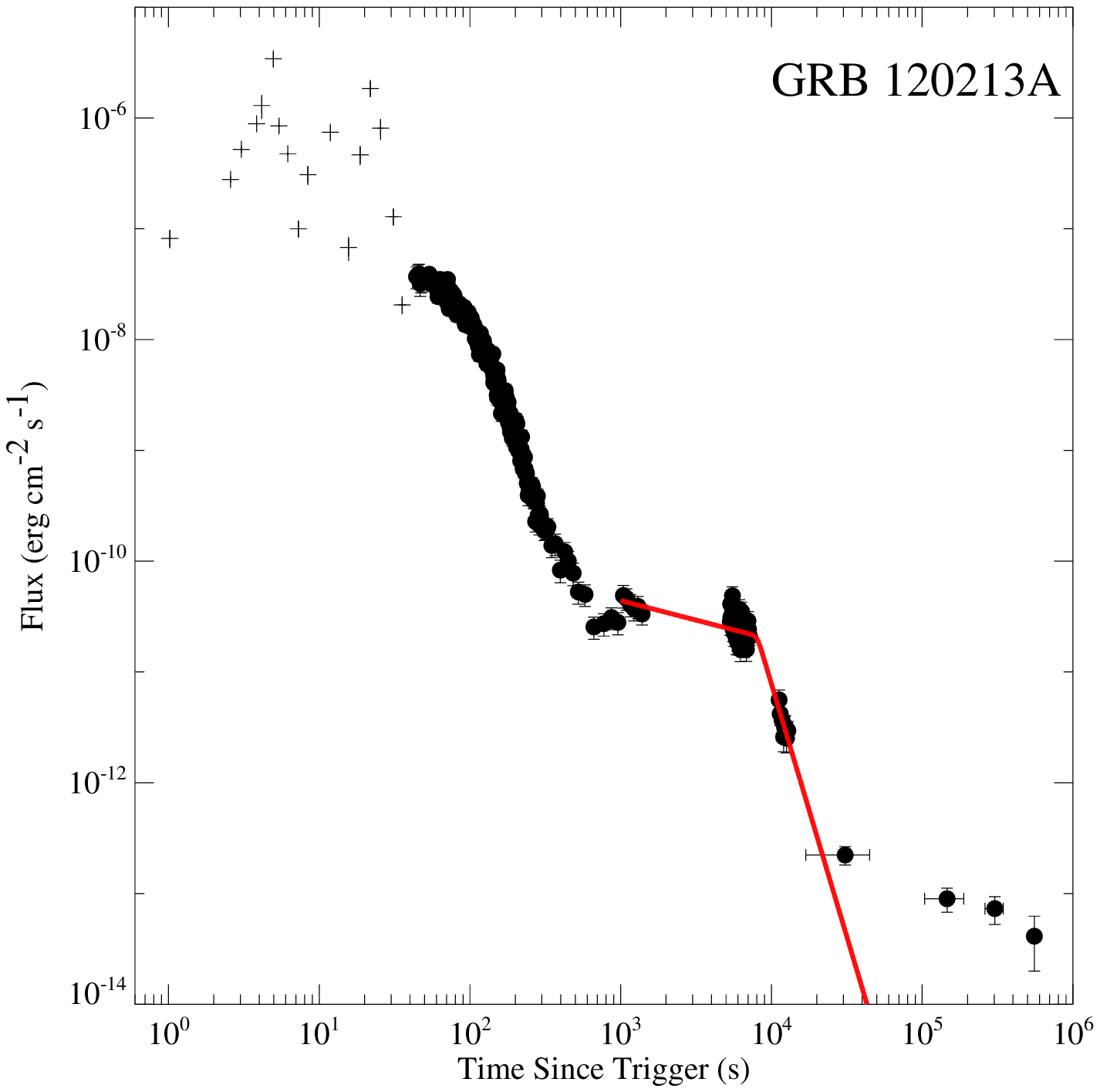}
\hfill
\includegraphics[angle=0,scale=0.35]{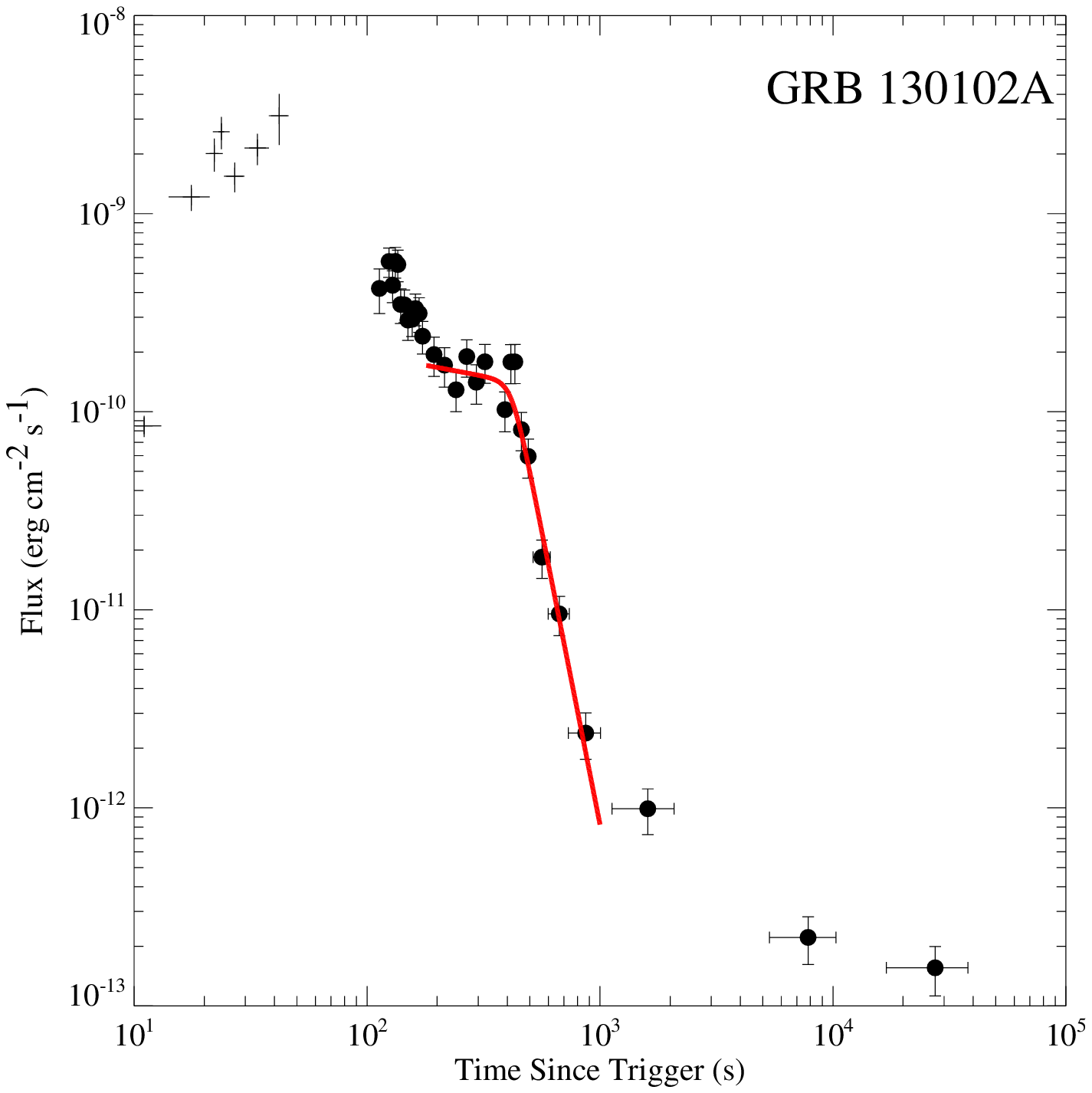}
\center\caption{The X-ray light curves of the GRBs in our Gold
sample. Plus signs are BAT data extrapolated to the XRT band,
and points (with error bars) are the XRT data. The red solid
curves are the best fits of the smooth broken power law model
to the data.}\label{X-ray}
\end{figure}

\begin{figure}
\includegraphics[angle=0,scale=0.35]{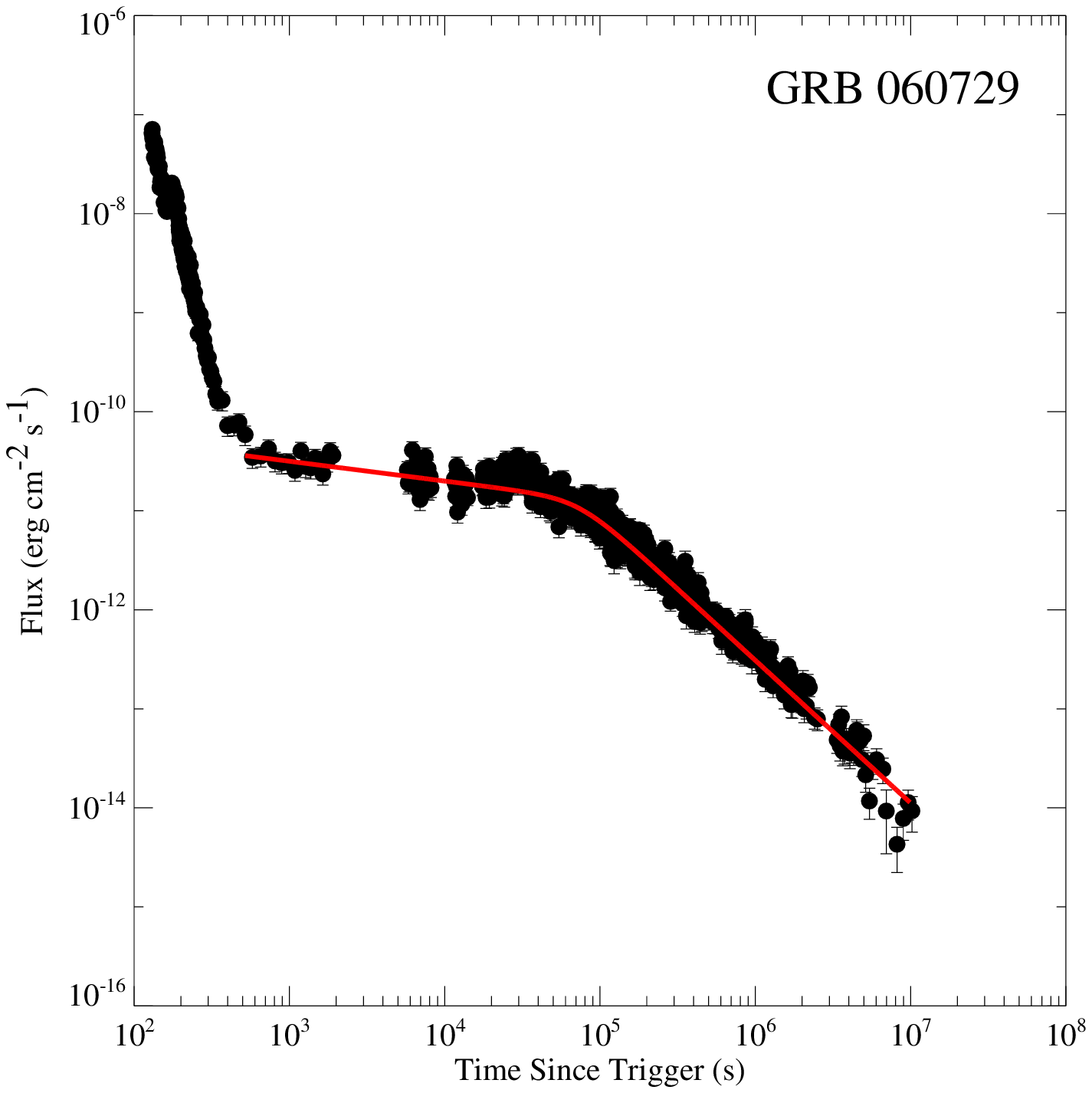}
\includegraphics[angle=0,scale=0.35]{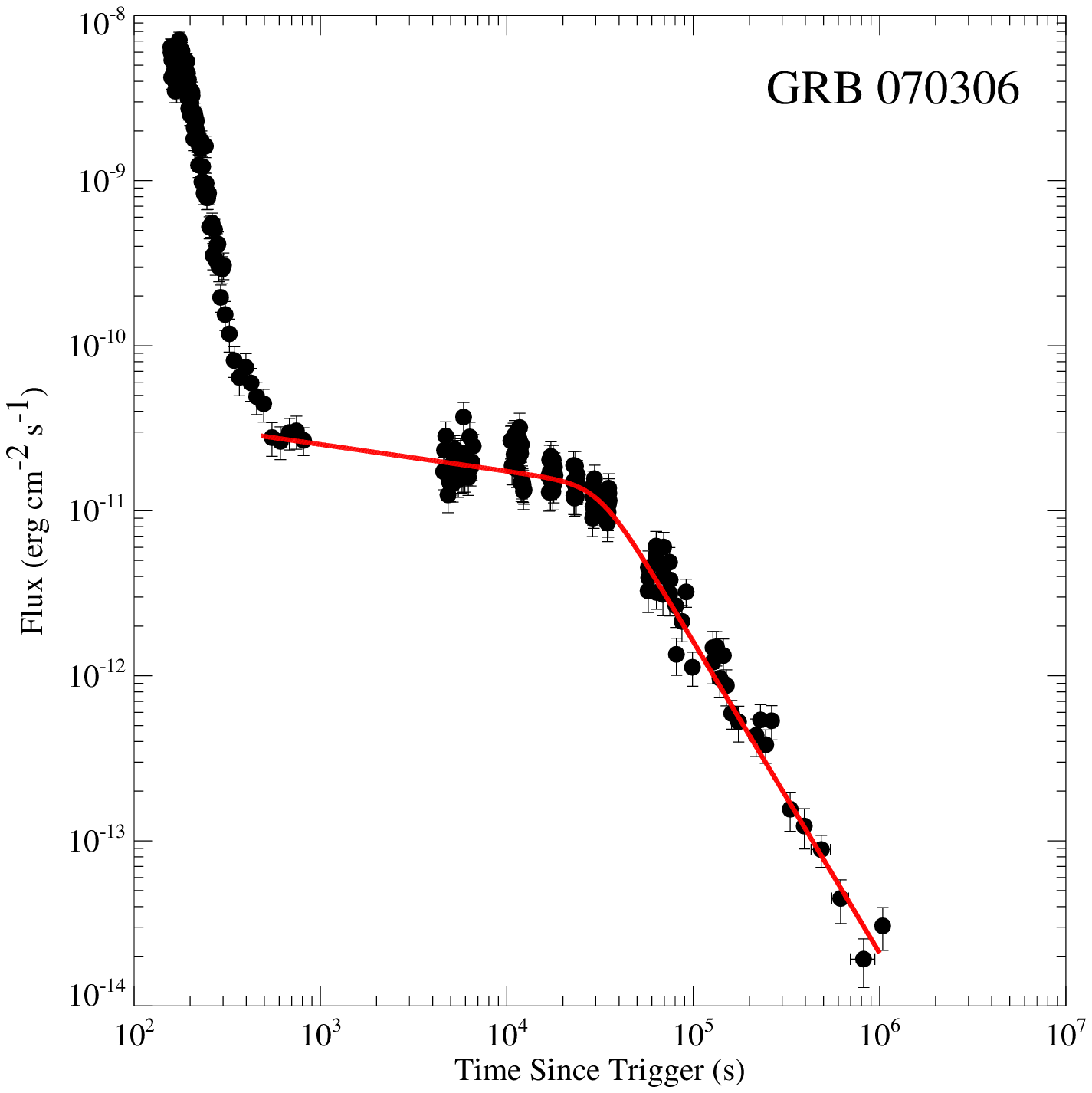}
\includegraphics[angle=0,scale=0.35]{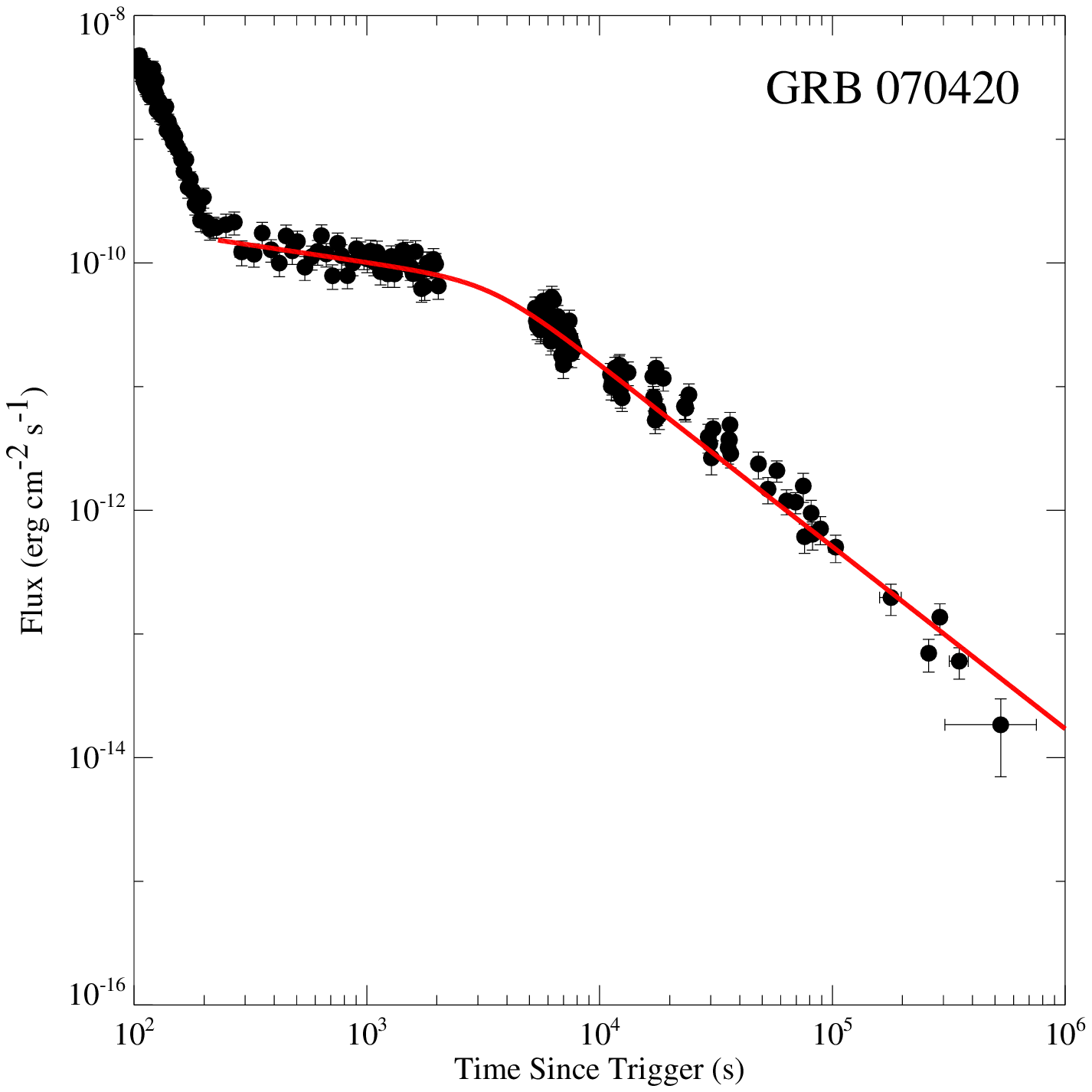}
\includegraphics[angle=0,scale=0.35]{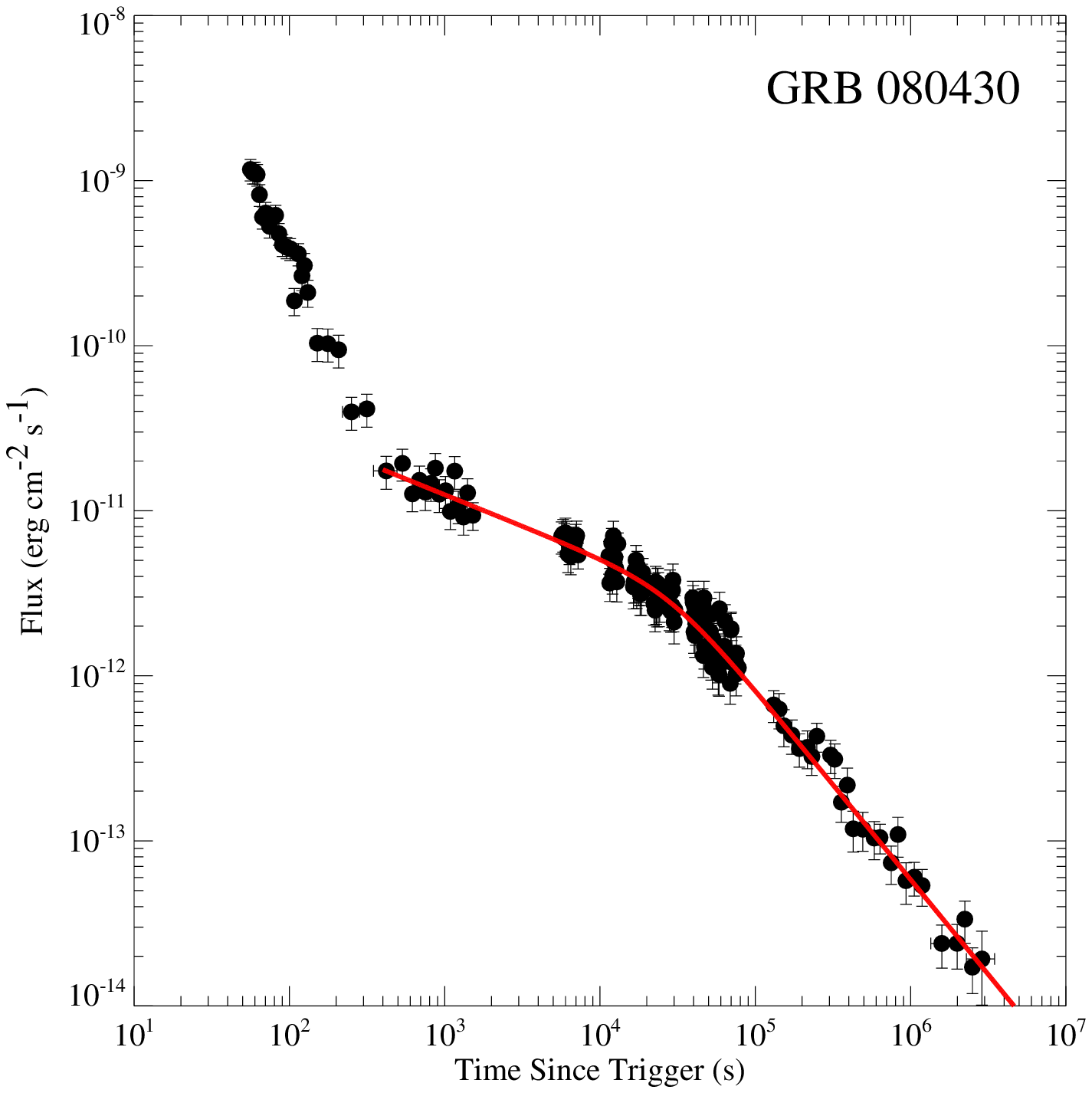}
\includegraphics[angle=0,scale=0.35]{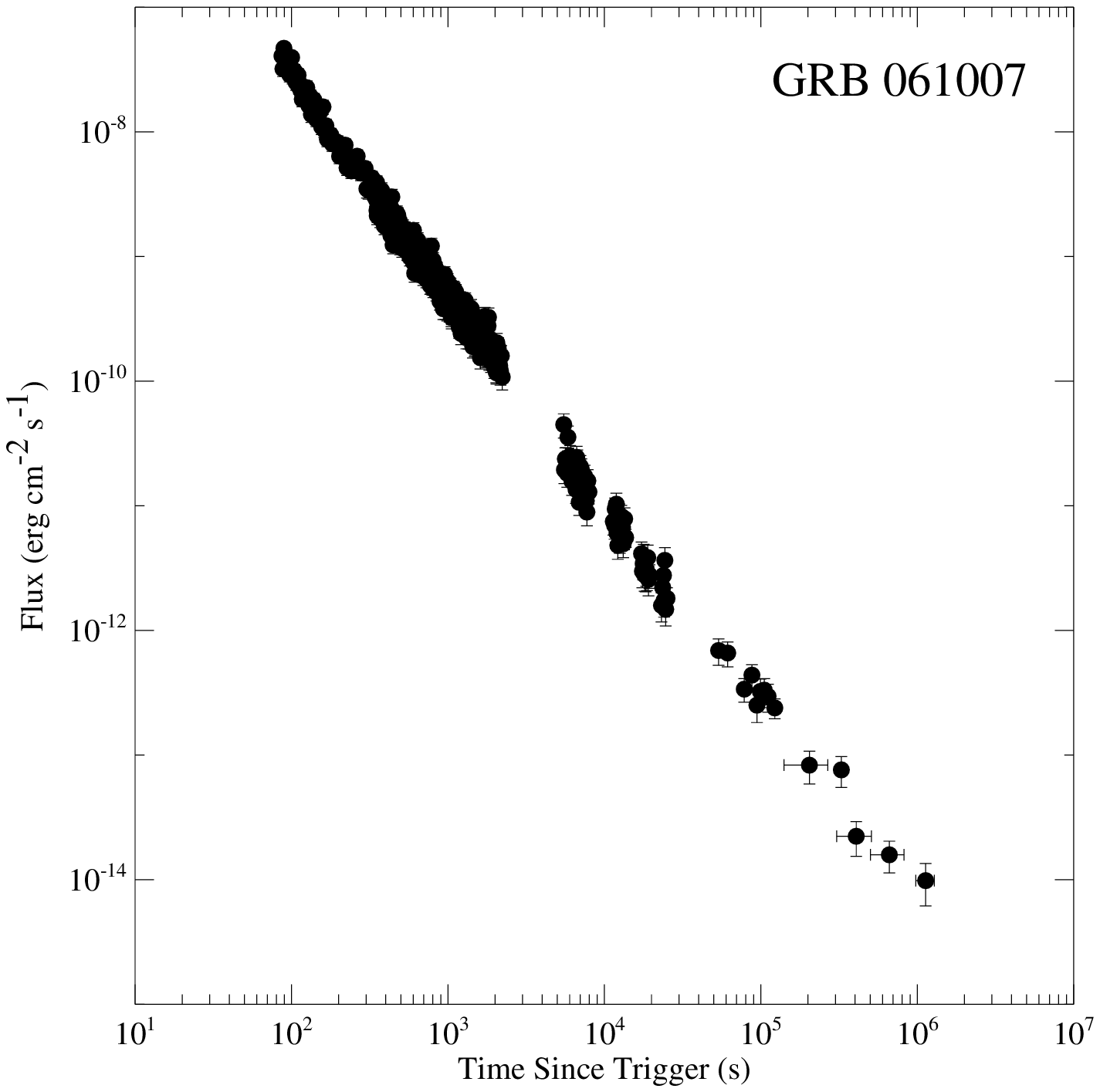}
\hfill
\includegraphics[angle=0,scale=0.35]{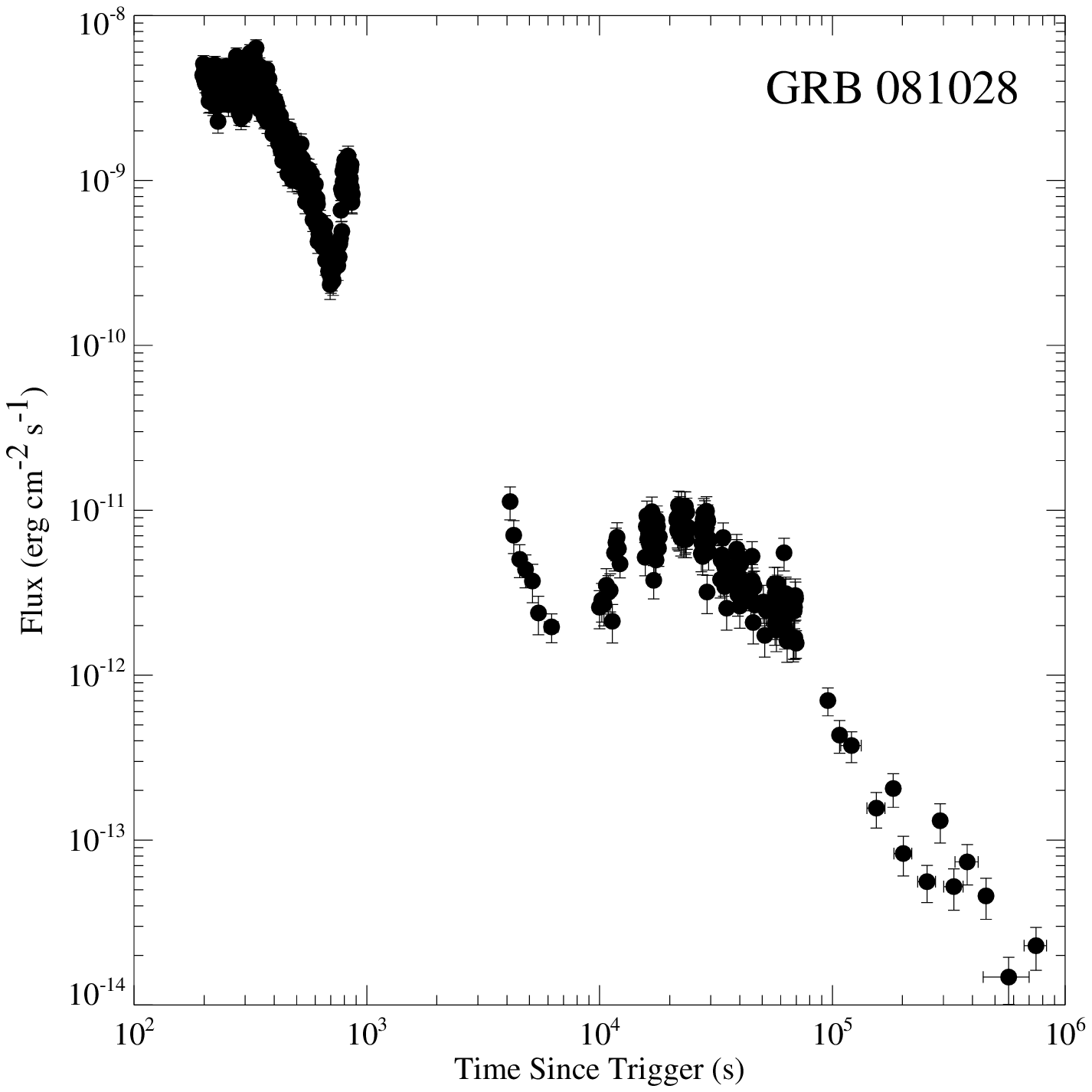}
\center\caption{Two cases of the X-ray light curves in our
Silver (GRB 060729 and 070306), Aluminum (GRB 070420 and 080430), and
Non-magnetar (GRB 061007 and 081028) sample. The red solid
curves are the best fits of the smooth broken power law model
to the data.}\label{X-ray2}
\end{figure}

\begin{figure}
\includegraphics[angle=0,scale=0.85]{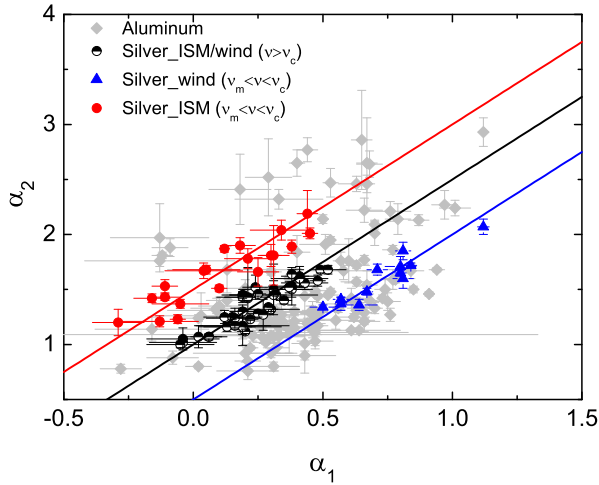}
\caption{ The temporal decay indices $\alpha_1$ vs. $\alpha_2$
for the ``Silver'' and ``Aluminum'' samples. The three solid
lines indicate the closure relations of three specific external
shock models invoking energy injection with the parameter $q=0$,
as is expected in the millisecond magnetar central engine model.
The colored data points belong to the Silver sample,  while grey
data points belong to the Aluminum sample.}
\end{figure}

\begin{figure}
\includegraphics[angle=0,scale=0.85]{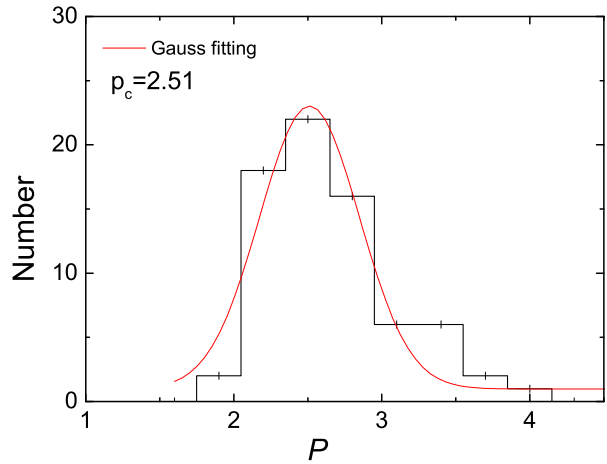}
\caption{ The distribution of electron spectral index $p$ derived
from the Silver sample. The solid line is the best Gaussian fit
with a center value $p_{c}=2.51$.}
\end{figure}

\begin{figure}
\includegraphics[angle=0,scale=0.85]{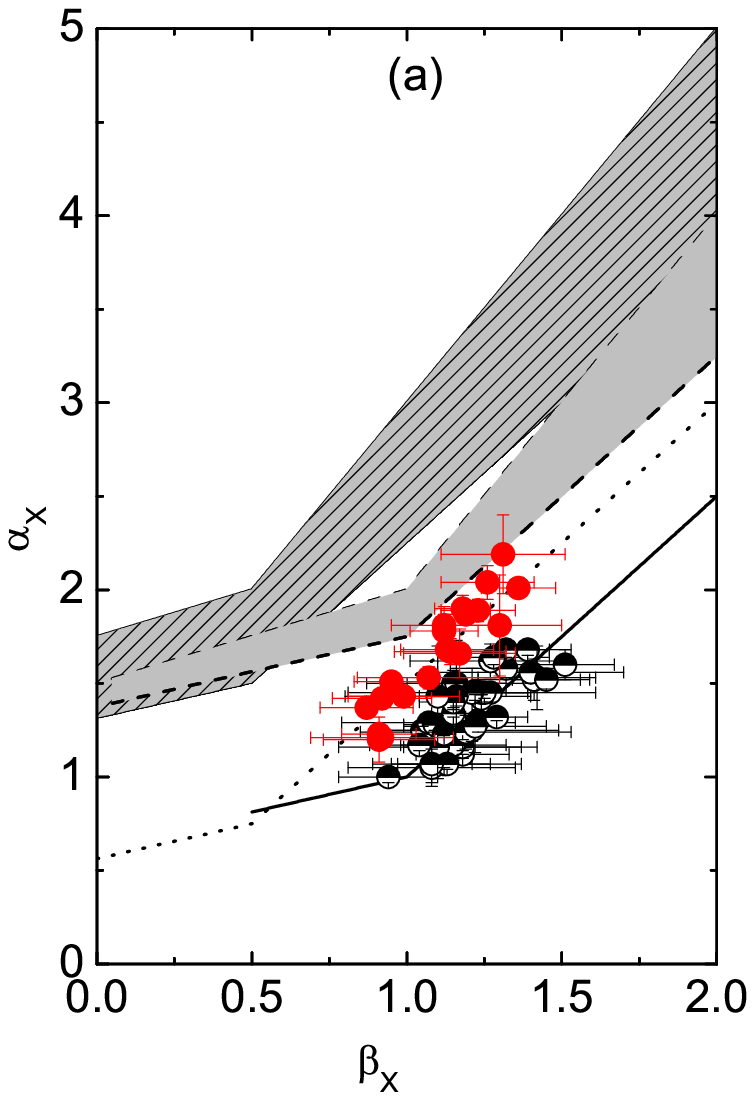}
\includegraphics[angle=0,scale=0.85]{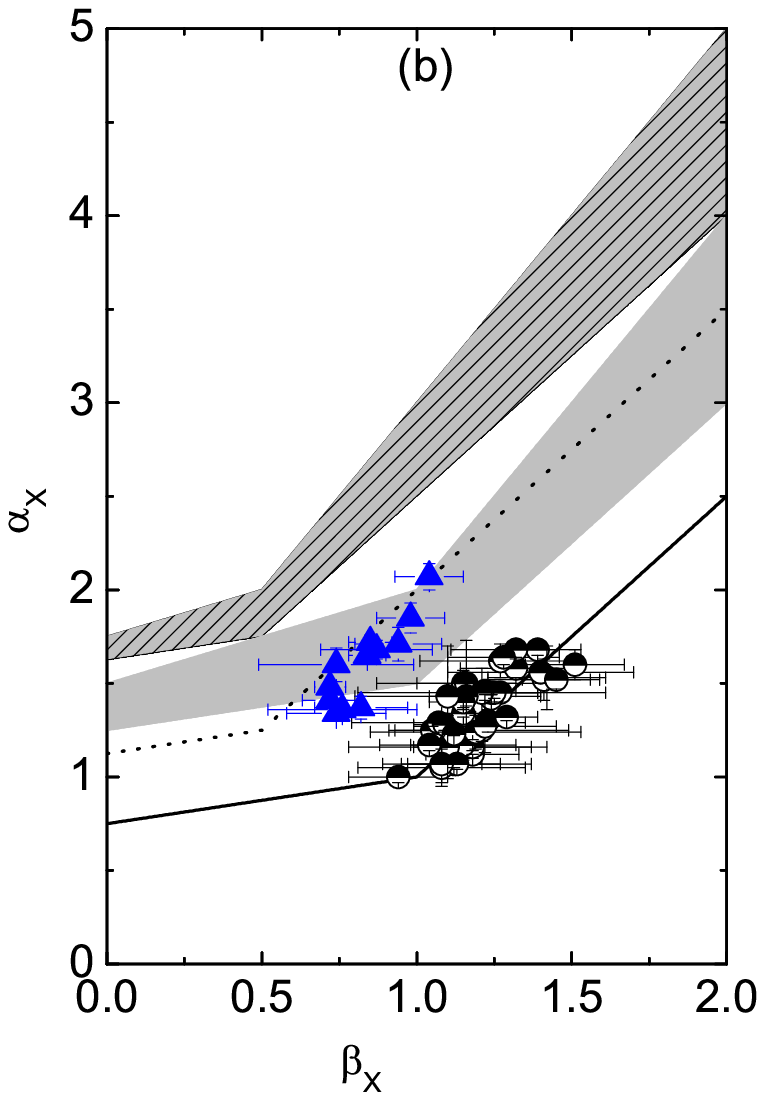}
\caption{The temporal decay index $\alpha$ against spectral index $\beta$
along with the closure relations of the external shock models
for the ``Silver'' sample. (a) The case of the ISM model: the solid
line (pre- jet break) and the shaded region (post jet break) are for
the spectral regime I ($\nu_{x}> \rm{max}(\nu_m,\nu_c)$), while the
dashed line (pre- jet break) and hatched region (post jet break) are
for the spectral regime II ($\nu_m<\nu_{x}<\nu_c$). Half-solid (black)
dots and solid (red) dots are for regime I and II, respectively.
(b) The case of the wind medium case. Same conventions, except that
triangles (blue) denote the spectral regime II.}
\end{figure}

\begin{figure}
\includegraphics[angle=0,scale=0.7]{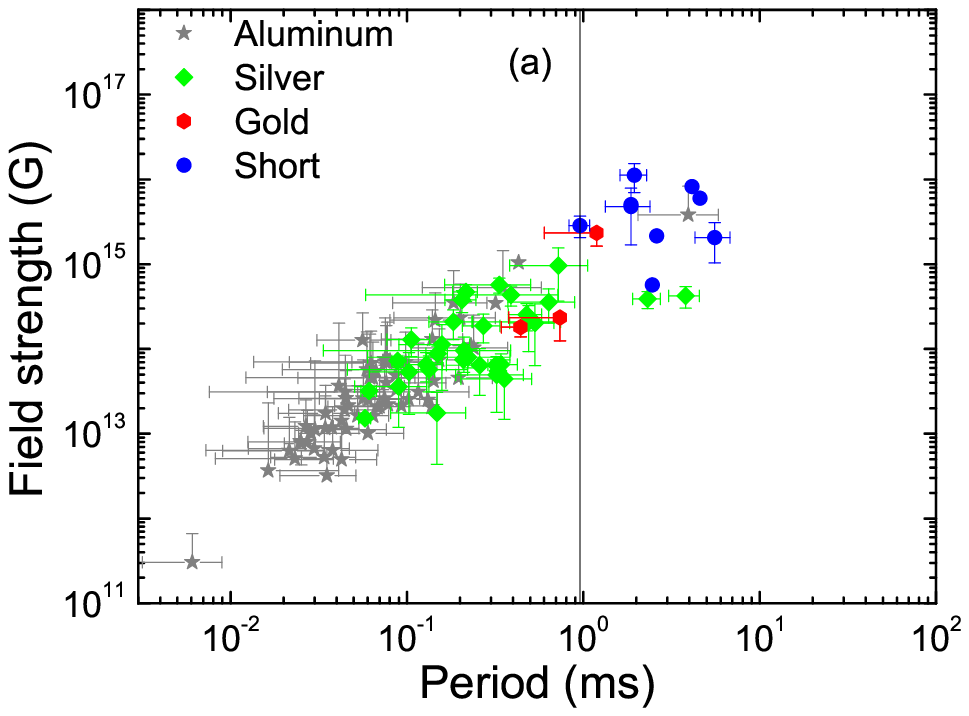}
\includegraphics[angle=0,scale=0.7]{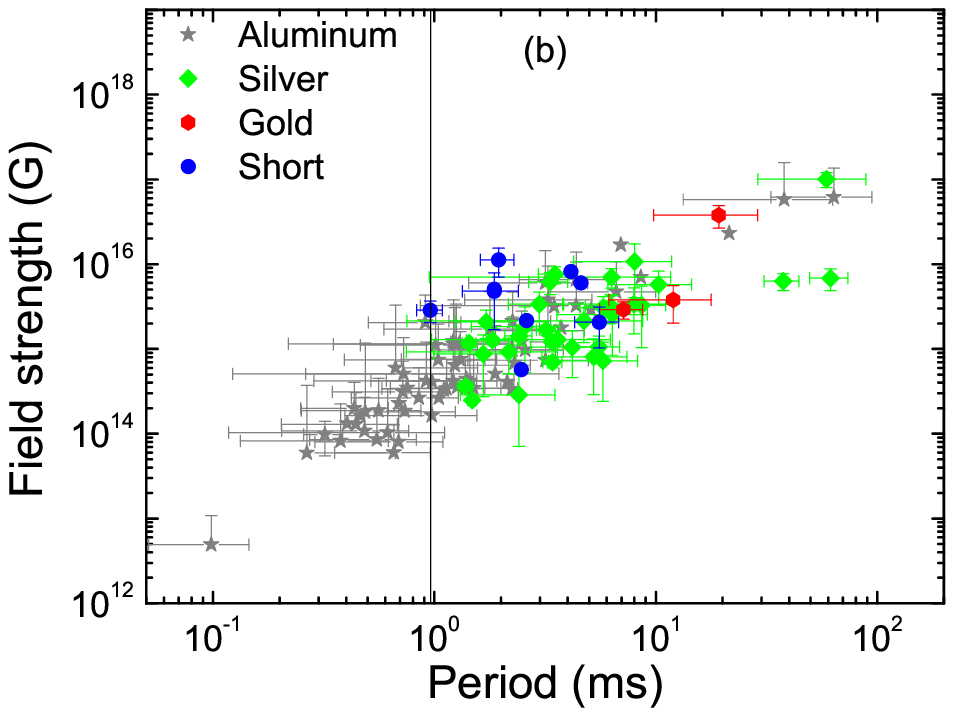}
\caption{The inferred magnetar parameters, initial spin period
$P_0$ vs. surface polar cap magnetic field strength $B_p$
derived for different magnetar samples: Gold (red hexagons),
Silver (green diamonds), Aluminum (grey), and short GRBs
(blue). (a) The case of isotropic winds; (b) The case with
beaming corrections. The vertical solid line is the breakup
spin-period for a neutron star (Lattimer \& Prakash 2004). }
\end{figure}

\begin{figure}
\includegraphics[angle=0,scale=0.7]{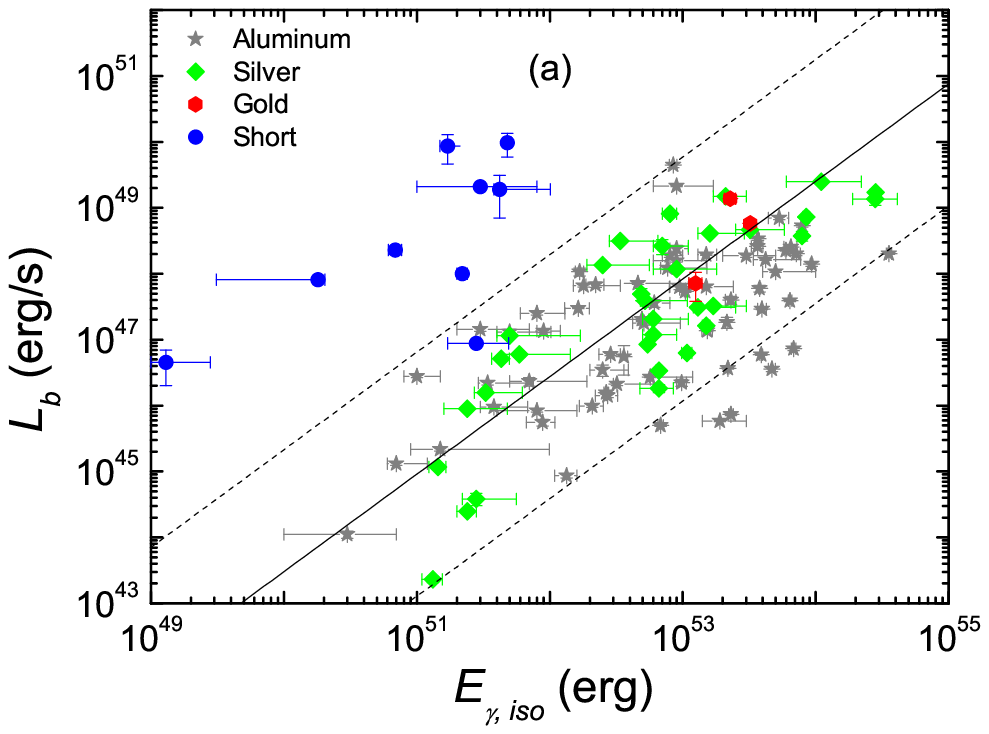}
\includegraphics[angle=0,scale=0.7]{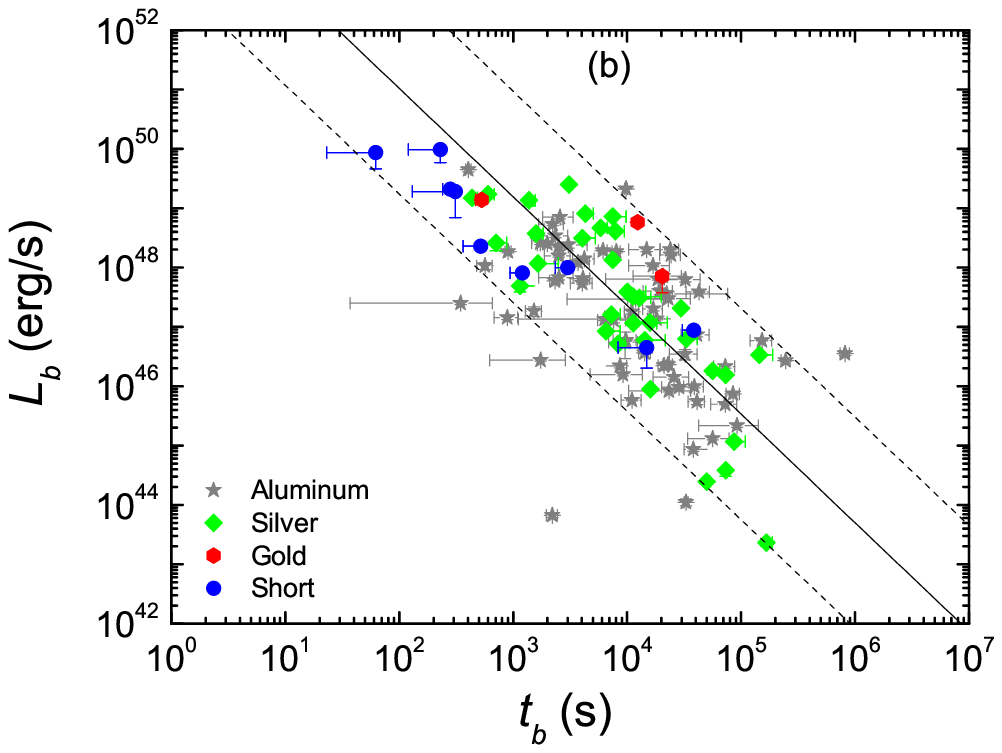}
\caption{The $L_{b}-E_{\gamma,iso}$ and
$L_{b}-t_{b}$ correlations for the GRBs in various magnetar
samples. The color convention is the same as Fig.5.
The solid line is a power-law fitting to the Gold and
Silver sample GRBs, and the two dashed lines denote the 2$\sigma$
region of the fits.}
\end{figure}

\begin{figure}
\includegraphics[angle=0,scale=0.7]{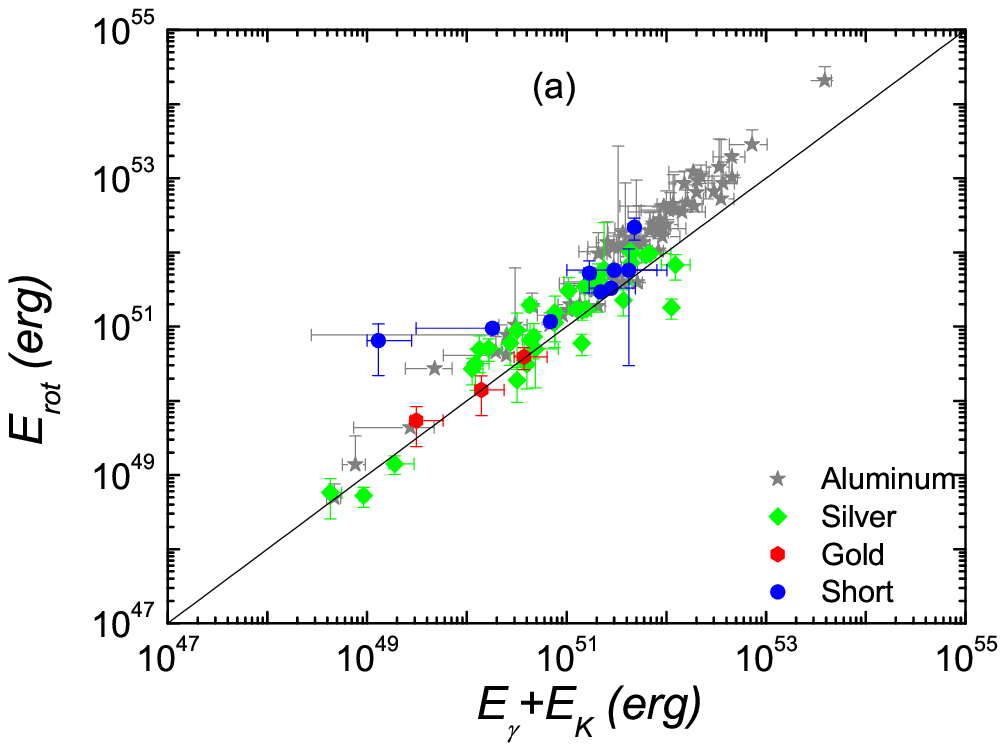}
\includegraphics[angle=0,scale=0.7]{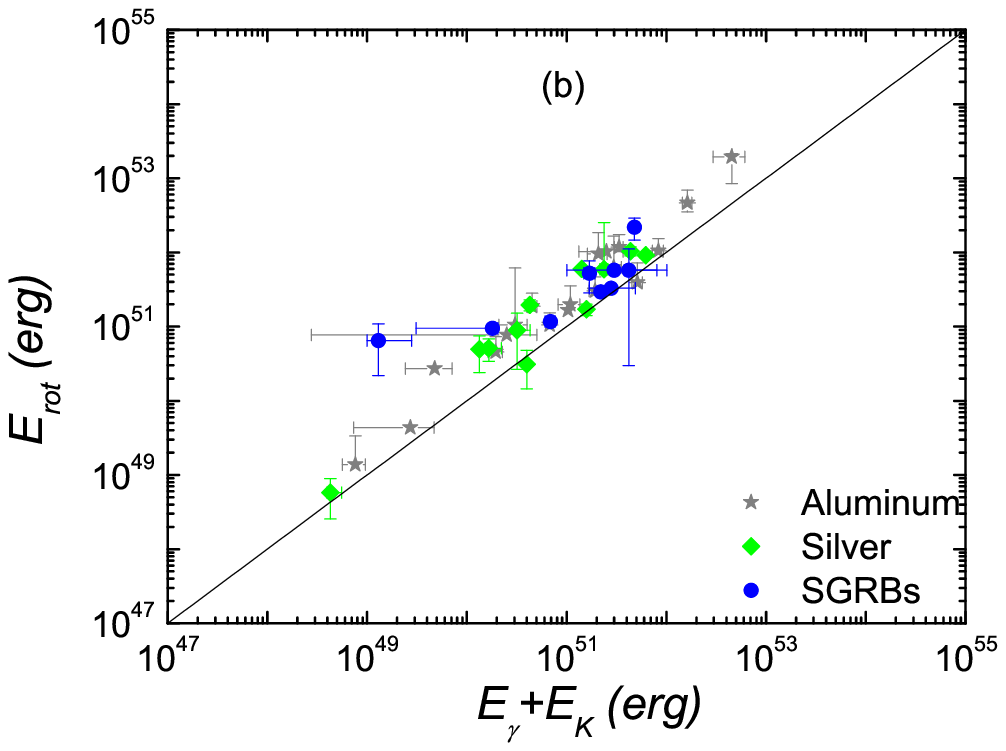}
\hfill\caption{A comparison between $(E_\gamma + E_{\rm K})$
and $E_{\rm rot}$. The color convention is the same as Fig.5.}
\end{figure}

\begin{figure}
\includegraphics[angle=0,scale=0.7]{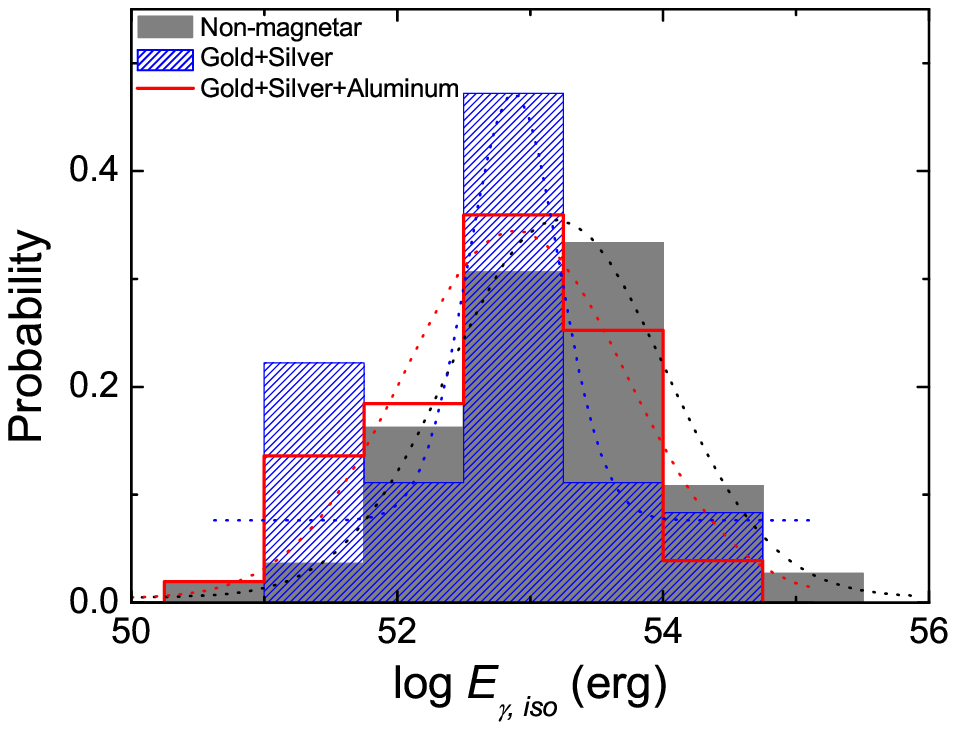}
\includegraphics[angle=0,scale=0.7]{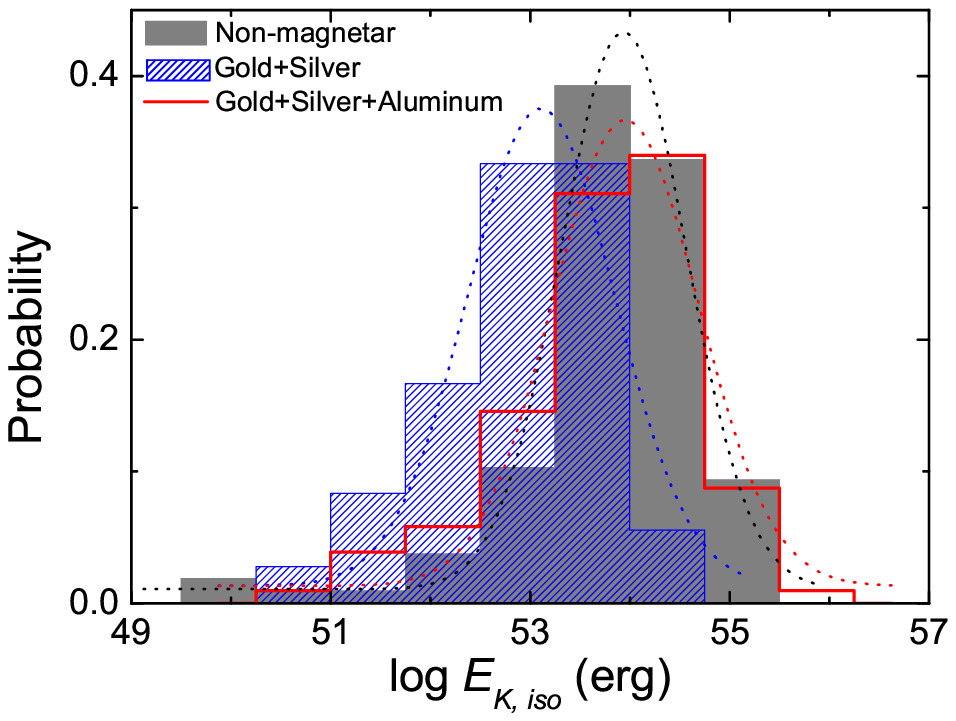}
\includegraphics[angle=0,scale=0.7]{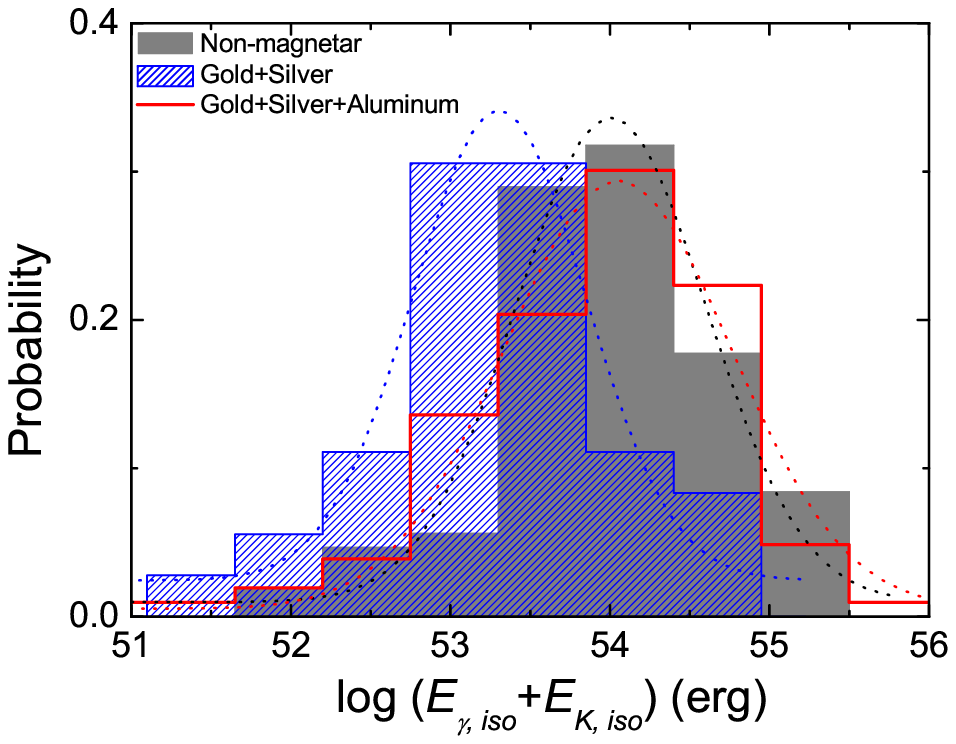}
\includegraphics[angle=0,scale=0.7]{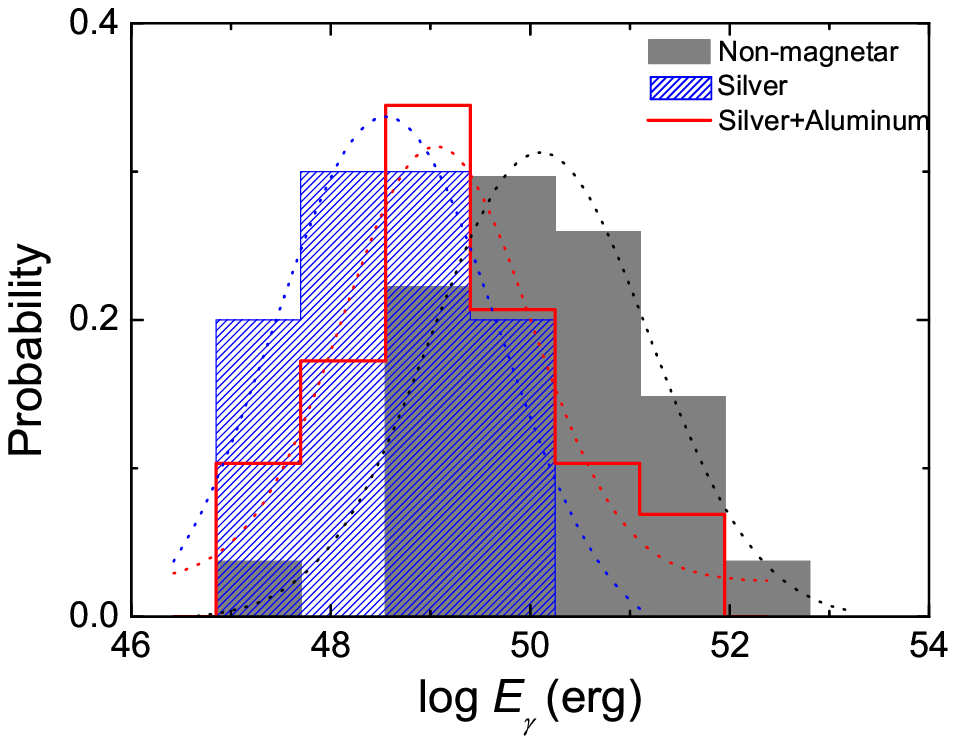}
\includegraphics[angle=0,scale=0.7]{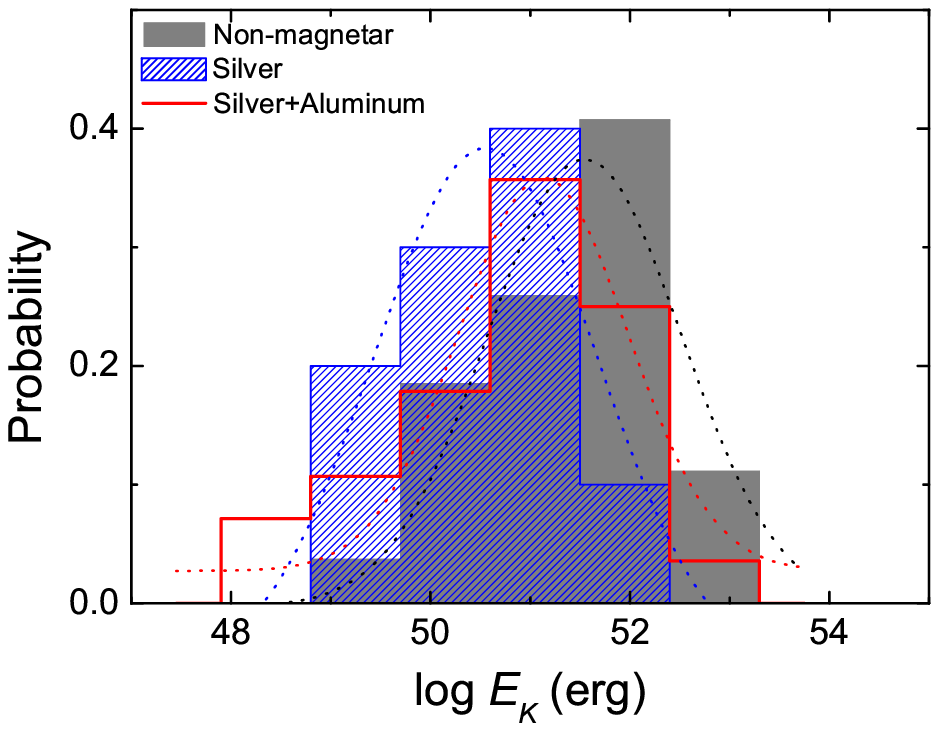}
\hfill
\includegraphics[angle=0,scale=0.7]{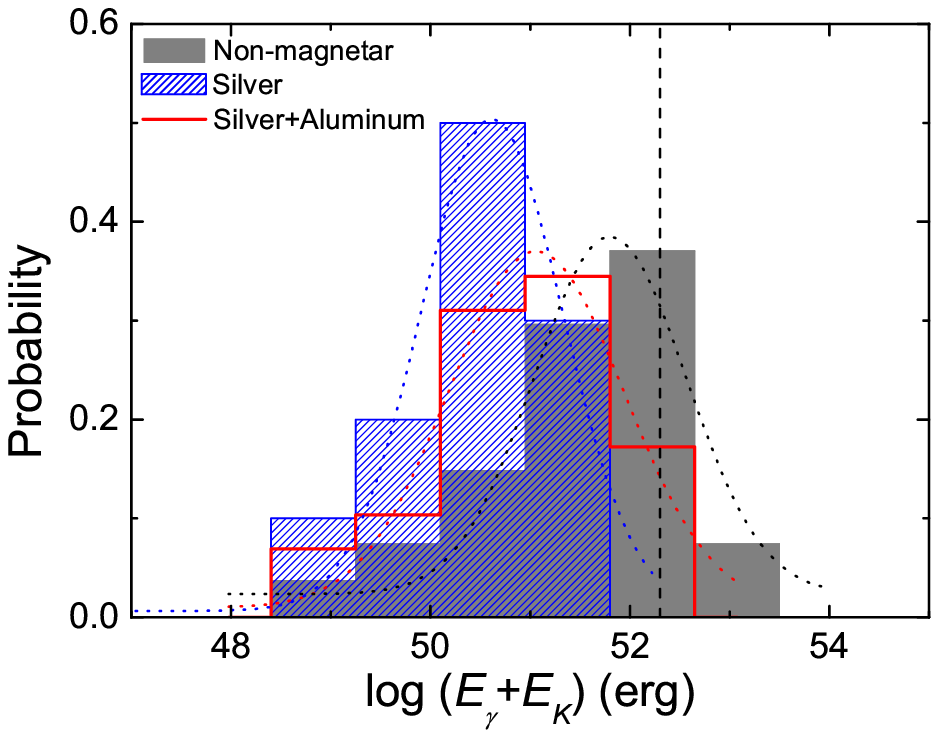}
\caption{Comparisons between the energy histograms of
the non-magnetar sample and the magnetar samples. The
non-magnetar, Gold+Silver, and Gold+Silver+Aluminum sample
histograms are denoted as grey filled, blue hatched, and red
open histograms, respectively. Best-fit Gaussian profiles are denoted in
black, blue, and red dotted lines, respectively. The six panels denote
histograms of $E_{\rm \gamma,iso}$, $E_{\rm K,iso}$, $(E_{\rm
\gamma,iso} +E_{\rm K,iso})$, $(E_{\gamma}$, $E_{\rm K}$, and
$(E_\gamma + E_{\rm K})$, respectively.}
\end{figure}

\begin{figure}
\includegraphics[angle=0,scale=0.7]{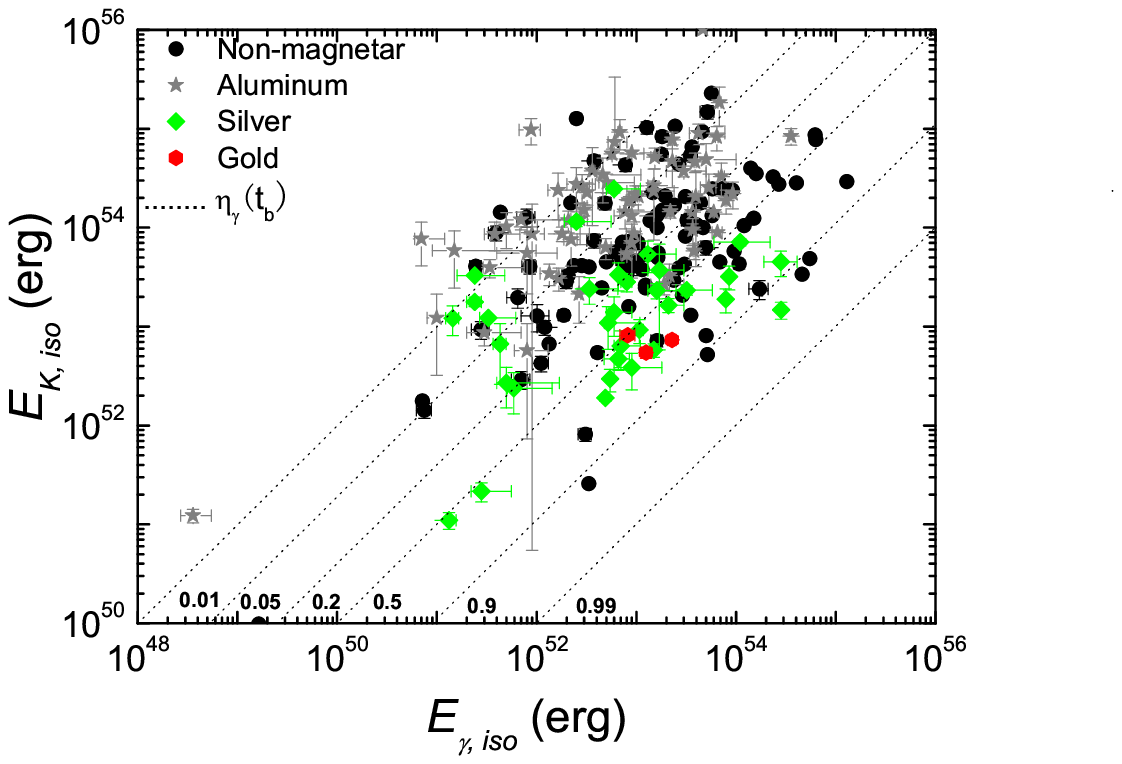}
\includegraphics[angle=0,scale=0.7]{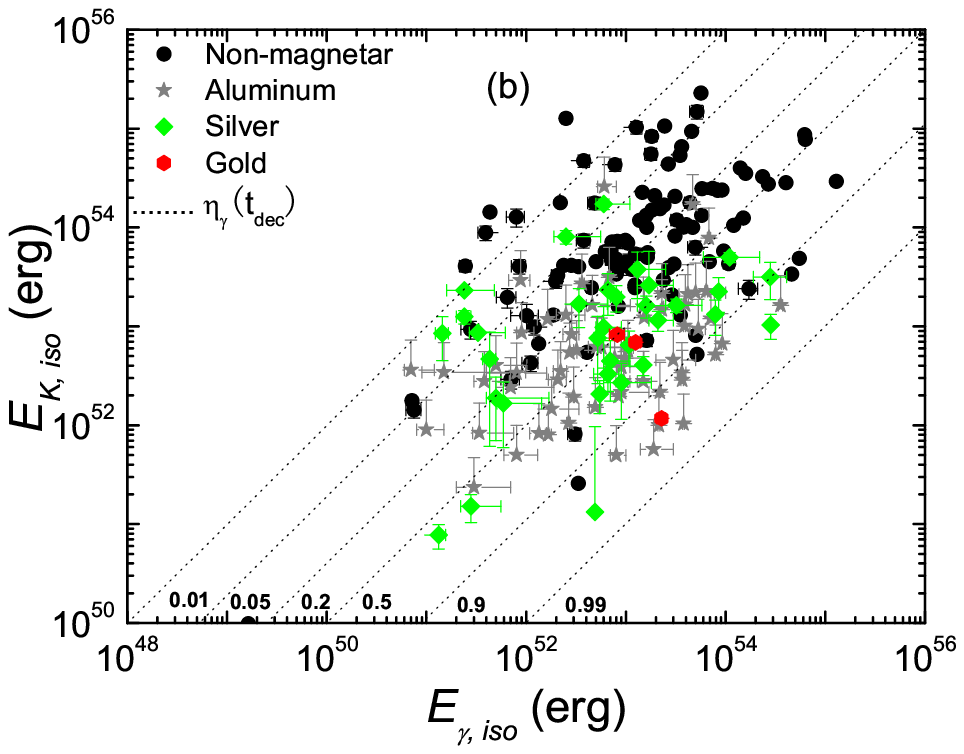}
\includegraphics[angle=0,scale=0.7]{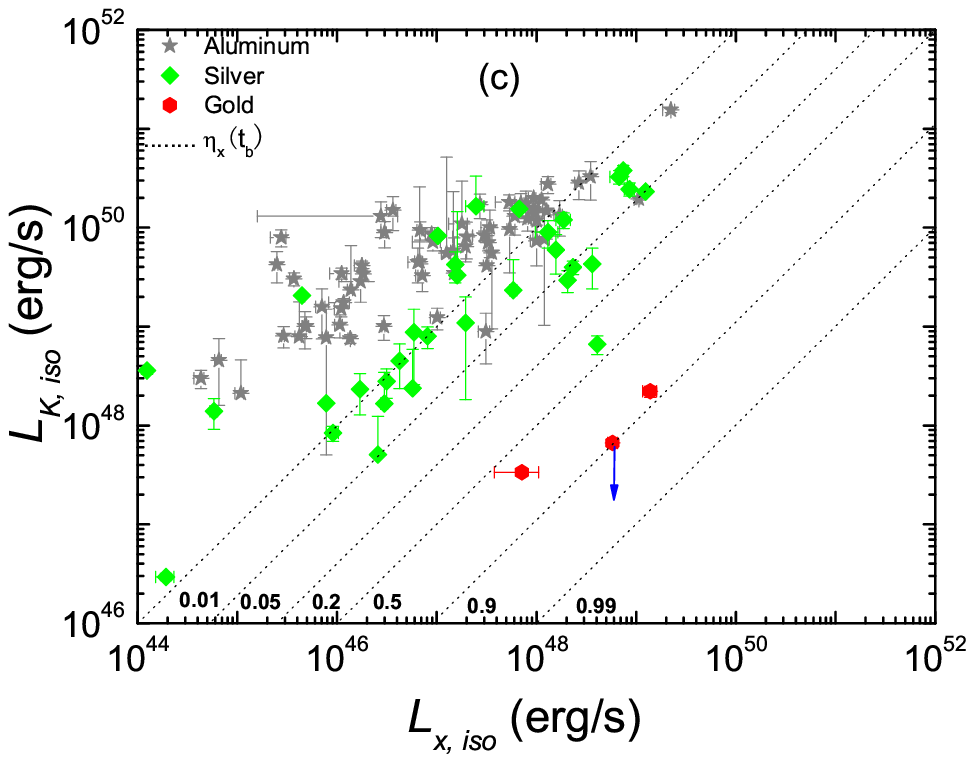}
\caption{(a) The $E_{\rm \gamma,iso} - E_{\rm K,iso}$ scattered
plot for all the GRBs with redshift measurements in our
samples: Gold (red), Silver (green), Aluminum (grey), and
non-magnetar (black). slanted dashed lines mark the constant
$\gamma$-ray efficiency ($\eta_\gamma$) lines. $E_{\rm K,iso}$
is calculated at $t_b$; (b) Same as (a), but with $E_{\rm
K,iso}$ calculated at $t_{dec}$; (c) The $L_{\rm X,iso} -
L_{\rm K,iso}$ scattered plot for the magnetar samples. Gold
(red), Silver (green), and Aluminum (grey). The constant X-ray
efficiency $\eta_{\rm X}$ lines are over plotted. The $L_{\rm
X,iso}$ value of silver and aluminum sample GRBs are all upper
limits. For one Gold sample GRB, $L_{\rm K,iso}$ is an upper
limit (denoted in the figure).}
\end{figure}

\begin{figure}
\includegraphics[angle=0,scale=0.7]{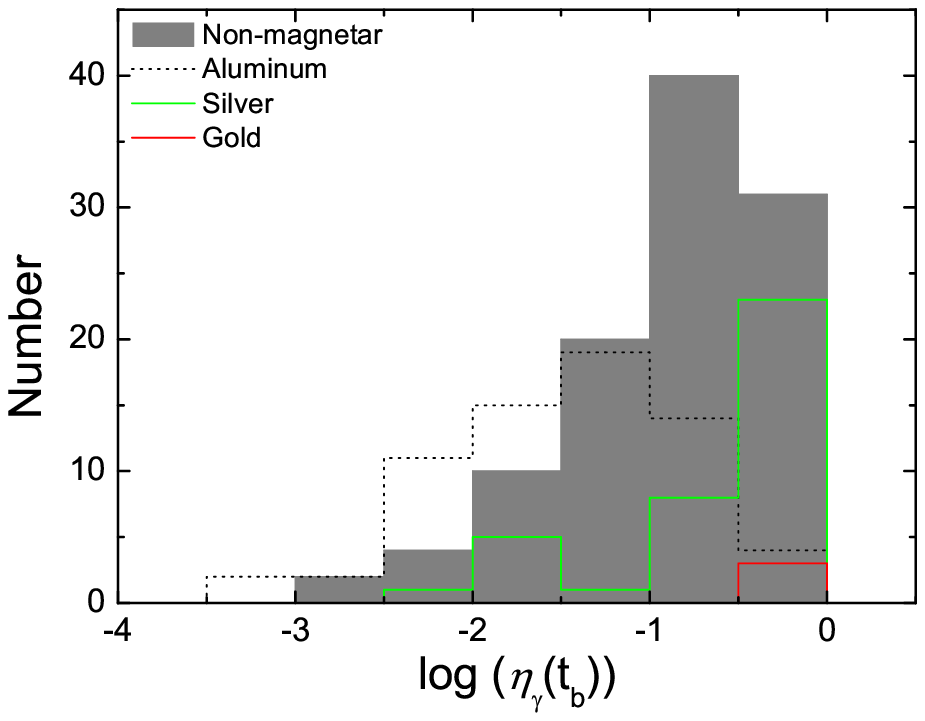}
\includegraphics[angle=0,scale=0.7]{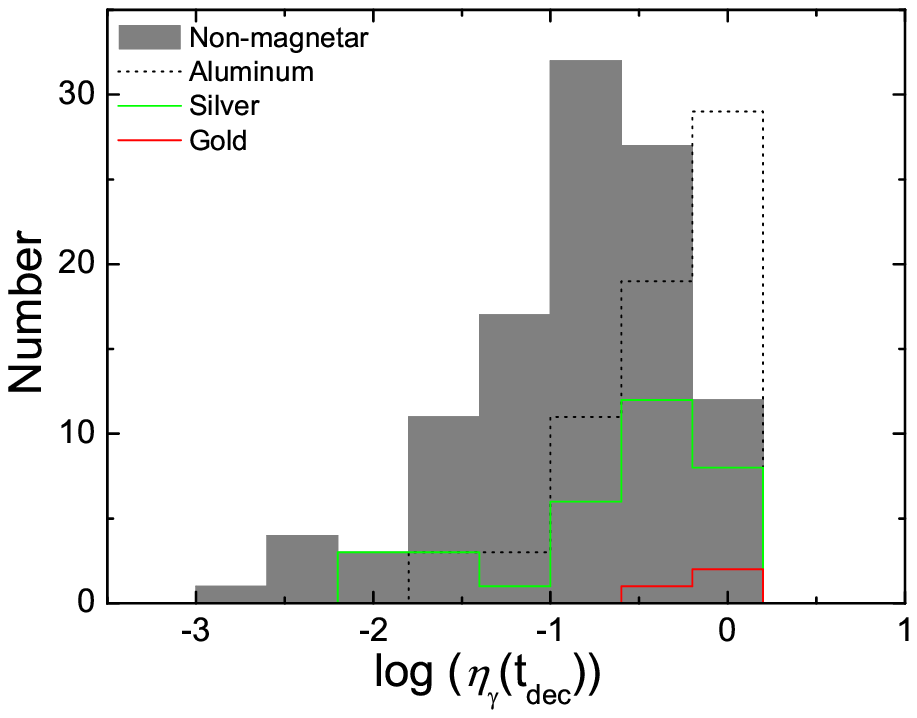}
\includegraphics[angle=0,scale=0.7]{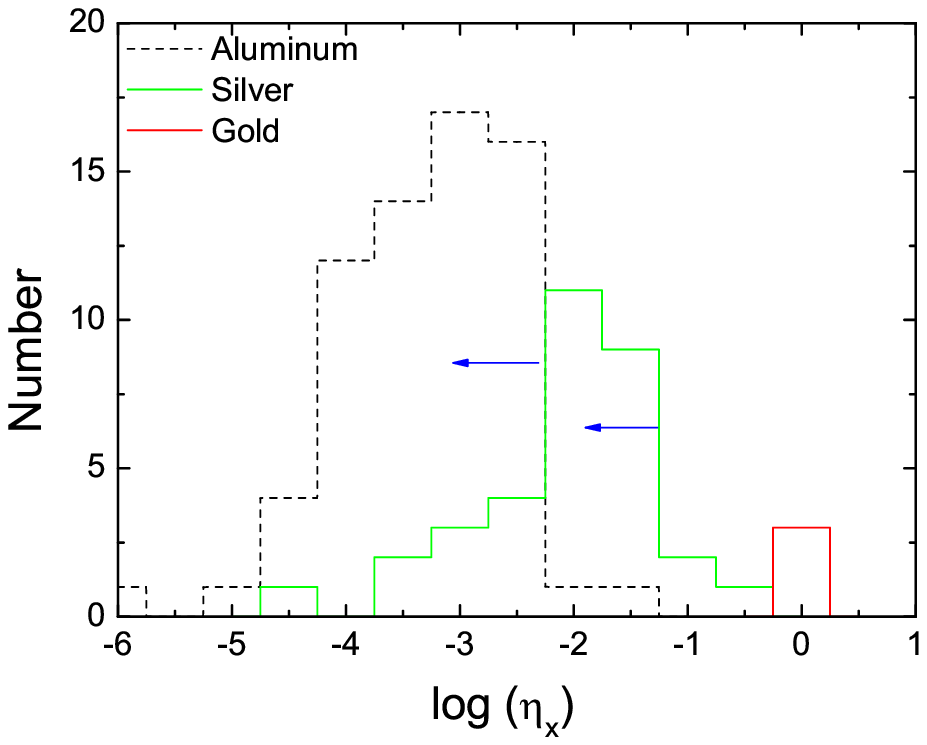}
\hfill
\caption{Histograms of $\eta_{\gamma}(t_b)$,
$\eta_{\gamma}(t_{dec})$ and $\eta_{\rm X}$ of our samples. For
$\eta_{\rm X}$, the silver and aluminum samples only give upper
limits.}
\end{figure}

\begin{figure}
\includegraphics[angle=0,scale=0.7]{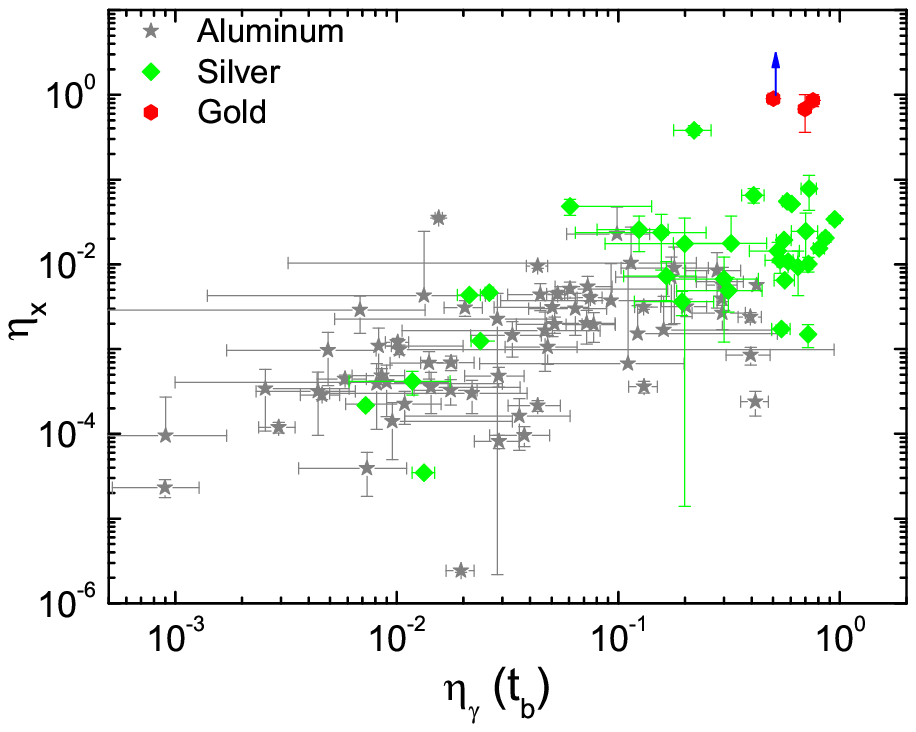}
\includegraphics[angle=0,scale=0.7]{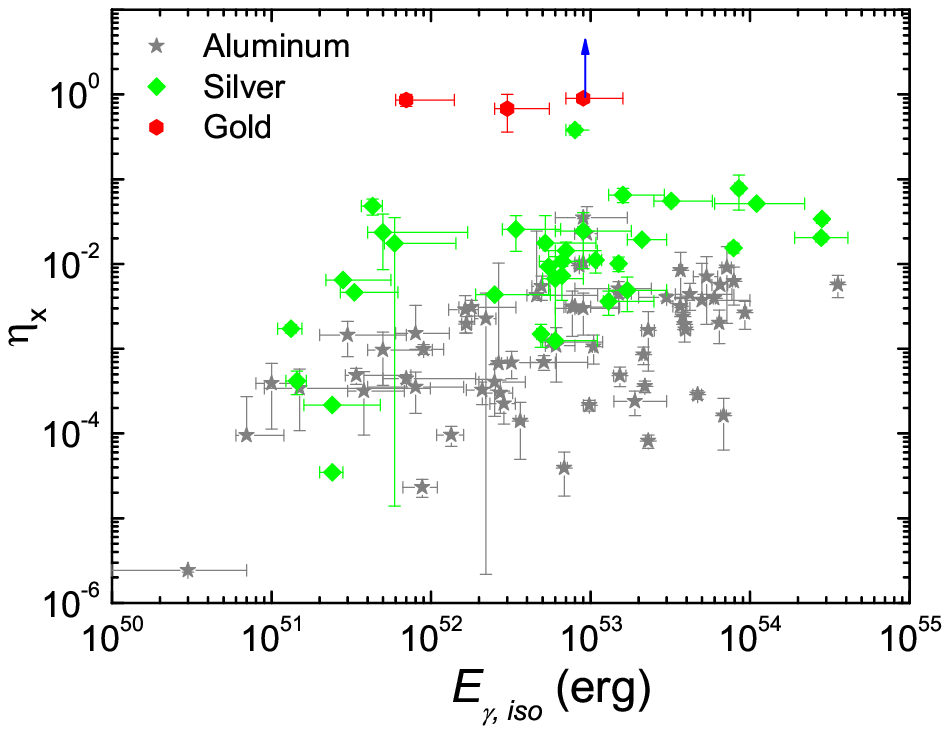}
\includegraphics[angle=0,scale=0.7]{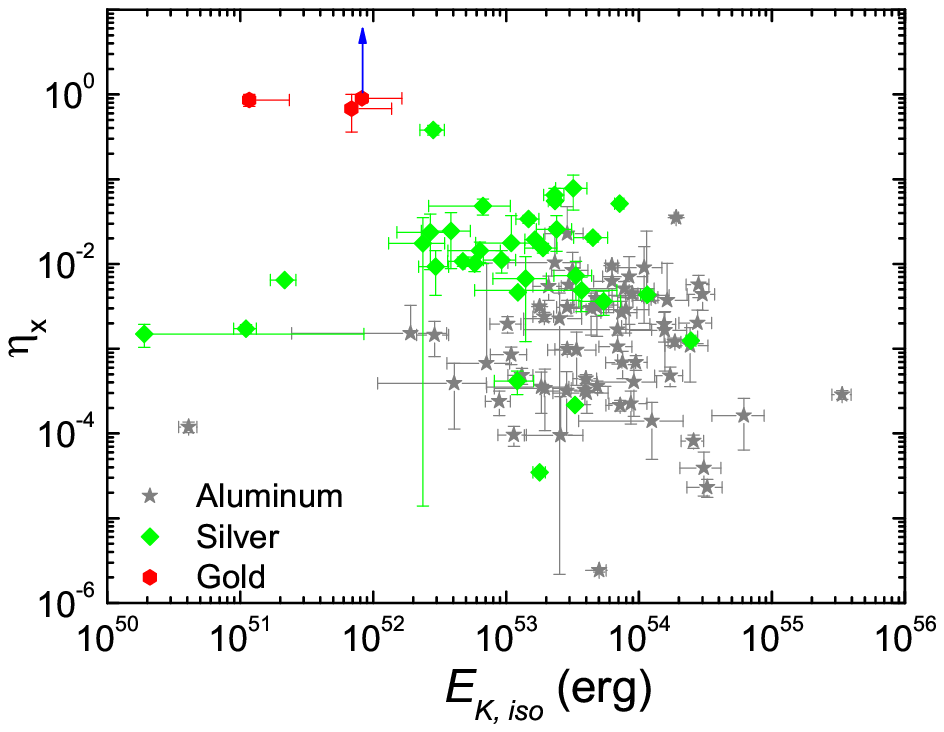}
\hfill
\includegraphics[angle=0,scale=0.7]{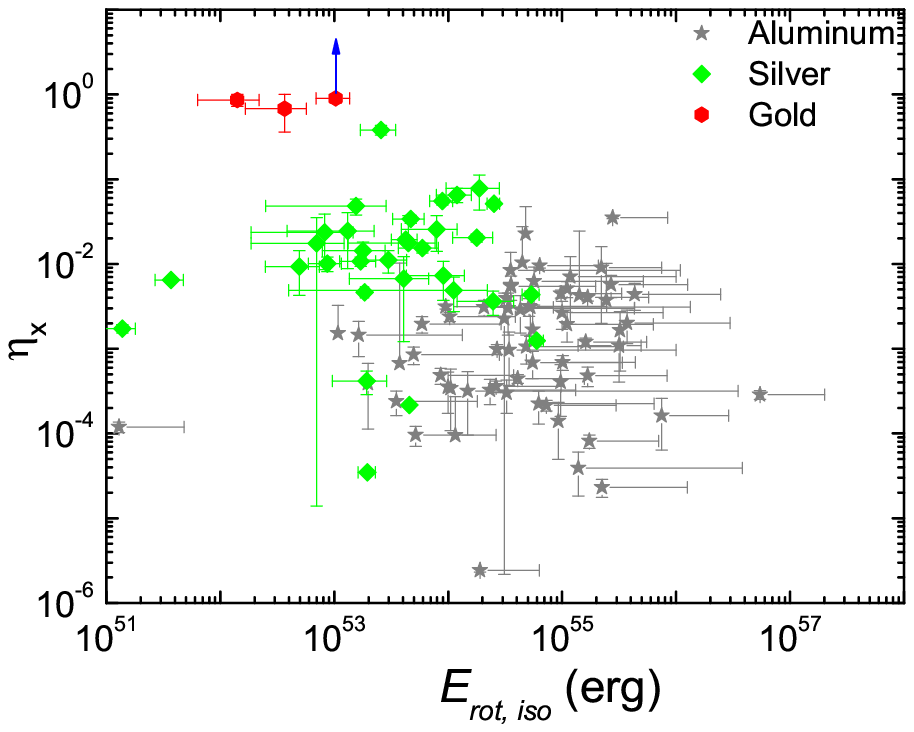}
\caption{The scatter plots of the X-ray efficiency $\eta_{\rm
X}$ vs. several parameters: $\eta_\gamma (t_b)$, $E_{\rm
\gamma,iso}$, $E_{\rm K,iso}$, and $E_{\rm rot}$. Color
conventions are the same as Fig.5. The $\eta_{\rm X}$ values of
all Silver and Aluminum sample GRBs are all upper limits. The
blue arrow shows the lower limit of one GRB in the Gold
sample.}
\end{figure}

\end{document}